\documentstyle[aps,preprint,epsfig]{revtex}
\tightenlines
\newcommand{\bea}{\begin{eqnarray}}
\newcommand{\eea}{\end{eqnarray}}
\newcommand{\beq}{\begin{equation}}
\newcommand{\eeq}{\end{equation}}
\newcommand{\kslash}{\mbox{$\not{\hspace{-0.6mm}k}$}}        
\newcommand{\nslash}{\mbox{$\not{\hspace{-0.6mm}n}$}}        
\newcommand{\vslash}{\mbox{$\not{\hspace{-1.03mm}v}$}}        
\newcommand{\pslash}{\mbox{$\not{\hspace{-1.03mm}p}$}}
\newcommand{\Dslash}{\mbox{$\not{\hspace{-1.03mm}D}$}}
\newcommand{\lslash}{\mbox{$\not{\hspace{-0.3mm}l}$}}
\newcommand{\qslash}{\mbox{$\not{\hspace{-1.03mm}q}$}}
\newcommand{\epsslash}{\mbox{$\not{\hspace{-0.5mm}\epsilon}$}}
\newcommand {\Q } {k_0}

\begin{document}
\preprint{\parbox{6cm}{\flushright CLNS 99/1644\\
LPT-Orsay-99-93\\
UCSD/PTH 99-17\\[1cm]}}
\title{Radiative leptonic decays of $B$ mesons in QCD\\[1cm]}
\author{Gregory P. Korchemsky$^1$, Dan Pirjol$^2$ and
Tung-Mow Yan$^3$\\[1cm]}
\address{$^1$ Laboratoire de Physique Th\'eorique,
Universit\'e de Paris XI, 91405 Orsay C\'edex, France}
\address{$^2$ Department of Physics,
University of California at San Diego, La Jolla, CA 92093}
\address{$^3$
Floyd R. Newman Laboratory of Nuclear
Studies, Cornell University, Ithaca, New York 14853}
\date{\today}
\maketitle

\begin{abstract}
We compute the form factors parametrizing radiative leptonic decays
of heavy mesons $B^+\to\gamma e^+\nu$ for photon energies much
larger than $\Lambda_{QCD}$, where perturbative QCD methods
for exclusive processes can be combined with the heavy quark
effective theory. The form factors can be reliably obtained in this
region in an expansion in powers of $\Lambda/E_\gamma$. The leading
term in this expansion displays an additional spin symmetry manifested
in the equality of form factors of vector and axial currents.
The leading twist form factors can be written as the convolution of
the $B$ meson light-cone wave function with a hard scattering amplitude,
which is explicitly calculated to one-loop order. The Sudakov
double logarithms of the form
$(\frac{\alpha_s}{\pi}\log^2\frac{2E_\gamma}{\Lambda})^n$
are resummed to all orders. As an application we present a
method for determining the CKM matrix element $|V_{ub}|$
from a comparison of photon spectra in $B$ and $D$ radiative
leptonic decays.
\end{abstract}

\pacs{pacs1,pacs2,pacs3}

\narrowtext
\section{Introduction}
The radiative leptonic decay $B^+\to \gamma\nu_\ell \ell^+$ has received a great deal of
attention in the literature \cite{BGW,Wyler,CdFN1,CdFN,AES,LCSR,LFM}, 
as a means of probing aspects of the
strong and weak interactions of a heavy quark system. The presence of the additional
photon in the final state can compensate for the helicity suppression of the rate
present in the purely leptonic mode. As a result, the branching ratio for the radiative leptonic
mode can be as large
as 10$^{-6}$ for the $\mu^+$ case \cite{LCSR}, which would open up a possibility for
directly measuring the decay constant $f_B$ \cite{CdFN}. A study of this decay can offer
also useful  information about the CKM matrix element $|V_{ub}|$.

Preliminary data from the CLEO collaboration indicate an upper
limit on the branching ratio ${\cal B}(B^+\to \gamma e^+\nu)$ of
$2.0\times 10^{-4}$ at the 90\% confidence level \cite{CLEOrad}. With better
statistics expected from the upcoming $B$ factories, the observation
and experimental study of this decay could become soon feasible.
It is therefore of some interest to have a good theoretical
control over the theoretical uncertainties affecting the relevant
matrix elements.

The hadronic matrix element responsible for this decay
can be parametrized in terms of two formfactors defined as
\bea\label{1}
& &\frac{1}{\sqrt{4\pi\alpha}}
\langle\gamma(p_\gamma,\epsilon) |\bar b\gamma_\mu (1-\gamma_5) q| B(v)\rangle =\\
& &\qquad\qquad \varepsilon(\mu,\epsilon^*,v,p_\gamma) f_V(E_\gamma) +
i[\epsilon_\mu^* (v\cdot p_\gamma) - (p_\gamma)_\mu (\epsilon^*\cdot v)]f_A(E_\gamma)\,.
\nonumber
\eea
The photon energy in the rest frame of the $B$ meson is $E_\gamma=v\cdot p_\gamma$.
The absolute normalization of the matrix element (\ref{1}) can be fixed, in the
limit of a soft photon, with the help of heavy hadron chiral perturbation theory.
In the limit of a massless final lepton, which we will consider everywhere in the
following, the leading contributions
to $f_V(E_\gamma)$ come from pole diagrams with a $J^P=1^-$ intermediate state, and
those to $f_A(E_\gamma)$ from $J^P=1^+$ states \cite{BGW}. The dominant contribution comes
from the $B^{*+}$ state, which is degenerate with $B^+$ in the heavy mass limit
\beq
f_V(E_\gamma) = \frac{Q_u\beta}{2(E_\gamma + \Delta)} f_{B^*} 
+ \cdots\,,
\eeq
where the ellipses stand for contributions from higher states with the same quantum
numbers and $\Delta = m_{B^*}-m_B$. 
$Q_q$ is the light quark electric charge in units of the electron 
charge. The hadronic parameter $\beta\simeq
3$ GeV$^{-1}$ parametrizes the $BB^*\gamma$ coupling in the
heavy quark limit \cite{beta}.
The formfactors $f_{V,A}$ have been also computed in the constituent quark
model \cite{CdFN,AES}, using light-cone QCD sum rules \cite{LCSR} and in a light-front
model \cite{LFM}.

Momentum conservation in the $B^+\to \gamma\nu_\ell \ell^+$ decay
can be written as $m_B v= q+p_\gamma$, with $q$ being the momentum of the lepton
pair. The photon energy is given by
\begin{equation}
E_\gamma=v\cdot p_\gamma=\frac{m_B}2 - \frac{q^2}{2m_B}
\end{equation}
and depending on the invariant mass of the lepton pair,
$0<q^2<m_B^2$ it takes values within the window $0< E_\gamma < m_B/2$.
In this paper we study the radiative leptonic decay $B^+\to 
\gamma\nu e^+$ in the kinematical region $\Lambda_{\rm QCD}\ll 
E_\gamma$
where the perturbative QCD methods developed in \cite{BL} for exclusive
processes can be applied. 

In the limit $m_b\to\infty$ and the photon energy satisfying 
$E_\gamma\gg \Lambda_{QCD}$,
the large momentum of heavy quark is carried away by the lepton pair
and does not affect the hadronic part of the decay. Therefore,
it becomes convenient to subtract away the large $m_b v$
component of the $b$ quark momentum and define a new transfer
momentum $\tilde q$ by
\begin{equation}
q = \tilde q + m_b v
\end{equation}
and $\tilde q_\mu={\cal O}(m_b^0)$. In the kinematical region
$E_\gamma \gg\Lambda_{QCD}$ this momentum is space-like $\tilde q^2 =
\bar\Lambda^2 - 2\bar\Lambda E_\gamma < 0$ with
$\bar\Lambda=m_B-m_b$ being the binding energy.

Introducing the subtracted momentum $\tilde q$ one notices
that in the leading $m_b\to\infty$ limit the kinematics of our
problem is very similar to the one
for the $\pi^0\to \gamma(p_\gamma)+\gamma^*(\tilde q)$ decay
discussed in \cite{BL}. This suggests to apply QCD factorization
theorems in order to expand the form factors $f_{V,A}(E_\gamma)$
in inverse powers of $\tilde q^2$, or equivalently $1/E_\gamma$.
In this formalism the form factors of interest can be written as the
convolution of a hard scattering amplitude with the transverse momentum
dependent wave function of the $B$ meson, $\psi(k_+,k_\perp)$.

We will work in a reference frame where the photon moves along the
``$-$'' light-cone direction and has light-cone components of the momentum
$ p_\gamma = (0,2E_\gamma/v_+, {\mathbf 0}_\perp)$.%
\footnote{Throughout the paper we shall use the following
definition of the light-cone components $k_\mu=(k_+,k_-,{\mathbf
k}_\perp)$ with $k_\pm=k_0\pm k_3$ and ${\mathbf
k}_\perp=(k_1,k_2)$.}
The transfer momentum is given by
$\tilde q=( \bar\Lambda v_+,  \bar\Lambda/
v_+ - 2E_\gamma/v_+,0_\perp)$. Then, to leading order in $1/E_\gamma$
(leading twist) and $\alpha_s$ we find
\bea\label{conv}
f_V(E_\gamma) = f_A(E_\gamma) = Q_q\sqrt{\frac{N_c}{2}}
\frac{1}{E_\gamma}\int
\frac{dk_+ d^2 k_\perp}{2(2\pi)^3} \frac{\psi(k_+,k_\perp)}{k_+}
+ {\cal O}(\Lambda^2/E_\gamma^2)\,.
\eea
The wave function $\psi(k_+,k_\perp)$ depends on the ``+'' light-cone
component, $k_+$, and transverse momentum, $k_\perp$, of the light quark
momentum in the $B$ meson.
Its properties are studied in Sec.~2, where its moments are related to
matrix elements of local heavy-light operators. The expression (\ref{conv})
for the form factors $f_{V,A}(E_\gamma)$ as integrals over the light-cone
wave function $\psi(k_+)$ is derived in Sec.~3. The radiative corrections
to this result induce a logarithmic dependence on $E_\gamma$, in addition
to the power law $1/E_\gamma$. These include doubly logarithmic Sudakov
corrections and mass-singular logarithms of the light quark
mass $\log(m)$, which are resummed in
Sec.~4. A few numerical estimates made with the help of a model wave function
are presented in Sec.~5, where we present also a method for extracting the
CKM matrix element $|V_{ub}|$ from a comparison of the photon spectra in
$B$ and $D$ radiative leptonic decays. A few details concerning the calculation
of the radiative corrections are presented in an Appendix.

\section{Light-cone $B$ meson wave function}

We consider a heavy $B$ meson with the flavor content $\bar bq$ and momentum
$p=m_B v$ moving along the $z$ axis. Its light-cone wave function can be expanded into
a sum of multiparticle Fock components
$|B\rangle = |\bar bq\rangle + |\bar bqg\rangle + \cdots$. The valence component
is written explicitly as
\bea\label{Bwf}
|B\rangle = \frac{\delta_{ab}}{\sqrt{N_c}}
\sum_{k_+,\vec k_\perp}
\psi(k_+, \vec k_\perp)
\frac{1}{\sqrt2}(|q^a(\tilde k,\uparrow)\bar b^b(\tilde k',\downarrow)\rangle -
|q^a(\tilde k,\downarrow)\bar b^b(\tilde k',\uparrow)\rangle)\,.
\eea
The light quark $q$ and heavy quark $\bar b$ in the $B$ meson have
light-cone momenta
$\tilde k\equiv (k_+,\vec k_\perp)$ and $m_b\tilde v + \tilde k'$,
respectively.  This gives the constraints
$k_+ + k'_+ = \bar\Lambda v_+$ and $\vec k_\perp + \vec k'_\perp = 0$, with
$\bar\Lambda = m_B-m_b$ the binding energy of the $B$ meson. The range of
variation of $k_+$ is the interval $(0, (m_b+\bar\Lambda) v_+)$,
corresponding to $k'_+ = (\bar\Lambda v_+, -m_b v_+)$.
The wave function $\psi(k_+)$  only takes values significantly different
from zero for $k_+/v_+ \stackrel{<}{\simeq} \bar\Lambda$.

The light-cone wave function $\psi(k_+,\vec k_\perp)$ is related to the usual
Bethe-Salpeter wave function $\Psi_{\alpha\beta}$ at equal light-cone ``time''
$\tau = x^0+x^3$
\bea
\Psi_{\alpha\beta}(k) = \int d^4\xi e^{ik\cdot \xi}
\langle 0|T^+\bar h_\beta(0) P(0,\xi) q_\alpha(\xi) |B(m_B v)\rangle
\eea
as
\bea\label{wf}
\Psi_{\alpha\beta}(k_+,\vec k_\perp) &\equiv&
\int_{-\infty}^\infty\frac{dk_-}{2\pi} \Psi_{\alpha\beta}(k) \\
&=&
\frac{\sqrt{N_c}}{\sqrt{(2m_b v_+)(2k_+)}}\cdot
\frac{1}{\sqrt2}\left(
u_\alpha(\tilde k,\uparrow) \bar v_\beta(v,\downarrow) -
u_\alpha(\tilde k,\downarrow) \bar v_\beta(v,\uparrow)
\right)\psi(k_+,\vec k_\perp)\,.\nonumber
\eea
The quark fields appearing in the definition of the Bethe-Salpeter wave function
are quantized on the light-cone
\bea
q_\alpha(x) = \sum_{\tilde k,\lambda}\frac{1}{\sqrt{2k_+}}
\left(a_{\tilde k,\lambda} u_\alpha(\tilde k,\lambda) e^{-ik\cdot x} +
b^\dagger_{\tilde k,\lambda} v_\alpha(\tilde k,\lambda) e^{ik\cdot x} \right)\,,
\eea
where the creation and annihilation operators satisfy $\{ a_{\tilde k,\lambda},
\,a^\dagger_{\tilde k',\lambda'}\} = \delta_{\tilde k,\tilde k'}\delta_{\lambda,\lambda'}$.
The light-cone spinors are defined as in \cite{pQCD} and are normalized according
to $\bar u(\tilde k,\lambda)\gamma_+ u(\tilde k,\lambda) = 2k_+$.

The static heavy antiquark field $h(x)$ is related to the usual $b$ field by
$b(x)=e^{im_b v\cdot x}h(x)$ and satisfies $\vslash h(x) = -h(x)$.
The path-ordered factor $P(0,\xi)=Pe^{-ig\int^0_\xi dz A(z)}$ is introduced
to ensure gauge invariance of the Bethe-Salpeter wave function.

Using the explicit expressions for the light-cone spinors $u(\tilde k,\lambda')$
and $\bar v(\tilde k',\lambda)$ given in \cite{BL} one finds the following
result for the wave function (\ref{wf}) in the limit of an infinitely heavy
$b$ quark
\beq
\Psi_{\alpha\beta}(k_+,\vec k_\perp) = -\frac{\sqrt{N_c}}
{\sqrt2 v_+ k_+}
\psi(k_+, k_\perp)
\left\{
(k_+ + \vec\alpha_\perp\cdot \vec k_\perp)
\Lambda_+ \frac{1+\vslash}{2}\gamma_5\right\}_{\alpha\beta}\,,
\eeq
satisfying the usual on-shell conditions
\beq
\kslash\Psi(k_+,\vec k_\perp) = 0\,,\qquad
\Psi(k_+,\vec k_\perp)\vslash = - \Psi(k_+,\vec k_\perp)\,.
\eeq
We denoted with
$\Lambda_+=\gamma_-\gamma_+/4$ the projector on the space of fast-moving
particles along the $+z$ axis.

It is convenient to define the one-dimensional wave
function\footnote{We denote both
$\psi(\tilde k)$ and $\psi(k_+)$ with the same letter. The distinction between
them is made through their arguments.} $\psi(k_+)$
by integrating over the transverse momenta
\beq
\psi(k_+) \equiv \int\frac{d^2 k_\perp}{(2\pi)^3}
\psi(k_+,\vec k_\perp)
\eeq
which satisfies
\bea
-\frac{1}{v_+}\sqrt{\frac{N_c}{2}}
\left\{
\Lambda_+ \frac{1+\vslash}{2}\gamma_5\right\}_{\alpha\beta}
\psi(k_+) =
\int_{-\infty}^\infty\frac{d\xi_-}{2\pi}
e^{ik_+\xi_-/2}
\langle 0|T\bar h_\beta(0) P(0,\xi) q_\alpha(\frac12\xi_- n_-) |B(p)\rangle\,.
\eea
Multiplying both sides with $(\gamma_+\gamma_5)_{\beta\alpha}$ gives
\beq\label{onedim}
\sqrt{2N_c} \psi(k_+) =
\int_{-\infty}^\infty\frac{d\xi_-}{2\pi}
e^{ik_+\xi_-/2}
\langle 0|T\bar h(0)\gamma_+\gamma_5 q(\frac12\xi_- n_-) |B(m_B v)\rangle\,.
\eeq
This relation can be used to express the moments of $\psi(k_+)$ in terms
of matrix elements of local operators. To see this,
the time-ordered product on the right-hand side of (\ref{onedim}) is expanded
into a power series of the separation on the light-cone. This gives
\bea
\sqrt{2N_c}
\psi(k_+)
= \sum_{n=0}^\infty \frac{1}{n!} \int \frac{d\xi_-}{2\pi}
\left(\frac{-i}{2}\xi_-\right)^n e^{ik_+\xi_-/2}
\langle 0|\bar h\gamma_+\gamma_5 (iD_+)^n q |B(m_B v)\rangle\,.
\eea
Taking the $j^{\rm th}$ moment with respect to $k_+$ one obtains the
desired connection to local operators (see also \cite{GrNe})
\bea\label{local}
\sqrt{N_c/2}
\int_0^\infty dk_+ (k_+)^j \psi(k_+) =
\langle 0|\bar h\gamma_+\gamma_5 (iD_+)^j q |B(m_B v)\rangle\,.
\eea

The first few moments of the wave function can be simply expressed in terms of
known hadronic quantities. For $j=0$ the corresponding matrix element on the
RHS of  (\ref{local}) is determined by the decay constant of
the $B$ meson in the static
limit defined as
\beq
\langle 0|\bar h\gamma_\mu\gamma_5  q |B(m_B v)\rangle =  f_B m_B v_\mu\,.
\eeq
One finds the normalization condition
\beq\label{normal}
\int_{0}^{\infty} \mbox{d}\left(\frac{k_+}{v_+}\right)
\psi\left(\frac{k_+}{v_+}\right) = \sqrt{\frac{2}{N_c}} f_B m_B \,.
\eeq

The first moment $j=1$ is given by the matrix element
\beq\label{n=1}
\int_0^\infty \mbox{d}\left(\frac{k_+}{v_+}\right)
\left(\frac{k_+}{v_+}\right) \psi\left(\frac{k_+}{v_+}\right) =
\sqrt{\frac{2}{N_c}}
\langle 0|\bar h\gamma_+\gamma_5 (iD_+) q |B(m_B v)\rangle =
\frac43
\sqrt{\frac{2}{N_c}} \bar\Lambda f_B m_B\,.
\eeq
This result agrees with the intuitive notion that the averaged spectator
quark momentum is proportional to the binding energy of the heavy hadron.
To prove it, one starts by writing the most general form for the
following matrix element, compatible with Lorentz covariance
\bea\label{D=1}
\langle 0|\bar h\gamma_\mu\gamma_5 (iD_\nu) q |B(m_B v)\rangle =
ag_{\mu\nu} + b v_\mu v_\nu \,.
\eea
The equation of motion for the light quark field $i\Dslash q(x)=0$ implies
the constraint $4a+b=0$. Another equation for these parameters can be obtained with
the help of the relation
\bea
\bar\Lambda v_\nu
\langle 0|\bar h\gamma_\mu \gamma_5 q |B(m_B v)\rangle =
\langle 0|\bar h\gamma_\mu \gamma_5  (\stackrel{\leftarrow}{iD_\nu}) q |B\rangle
+ \langle 0|\bar h\gamma_\mu \gamma_5  (\stackrel{\rightarrow}{iD_\nu}) q |B\rangle\,.
\eea
Multiplying both sides with $v^\nu$ and using the static quark equation of motion
$iv\cdot D h(x)=0$, one obtains $a+b= \sqrt2 \bar\Lambda f_B m_B$.
Solving for $a$ and $b$ gives the result presented in (\ref{n=1}).

In the presence of radiative corrections, the connection between the
light-cone wave function and matrix elements of local operators is changed.
For example, the zeroth moment (\ref{normal}) acquires a scale dependence
typical of matrix elements of operators in the effective theory with heavy
quarks \cite{VoSh,PoWi}. In the rest frame of the $B$ meson $(v_+=1)$ this
is given by, after $\overline{MS}$ renormalization, 
\bea\label{normal1}
& &\sqrt{\frac{2}{N_c}} f_B(\mu_{\rm h}) m_B =\\
& &\qquad
\int_0^\infty \mbox{d}k_+ \int\frac{d^2 \vec k_\perp}{(2\pi)^3}
\psi(k_+,k_\perp)\left(1 + \frac{\alpha_s C_F}{4\pi}
\left\{
-\frac32\log\frac{m^2}{\mu_{\rm h}^2} + \Delta_{IR}(\Q)
+ F(\Q) \right\}\right)\nonumber\,. \eea The term $\Delta_{IR}(\Q )$
contains an IR singularity, which is regulated with dimensional
regularization in $D=4-2\epsilon$ dimensions. The quantities
$\Delta_{IR}(\Q )$ and $F(\Q )$  are given by 
\bea\label{JIR}
\Delta_{IR}(\Q ) &=& 2
\left(N_\varepsilon^{IR}+\log\frac{\mu^2}{m^2}\right)
\left[\frac{\Q }{2\sqrt{\Q ^2-m^2}}
\left(\log\frac{\Q +\sqrt{\Q ^2-m^2}}{\Q -\sqrt{\Q ^2-m^2}} - 2\pi
i\right) -1\right]\\ &-&  \frac{\Q }{\sqrt{\Q ^2-m^2}}\left[
\mbox{Li}_2\left( \frac{2\sqrt{\Q ^2-m^2}}{\Q +\sqrt{\Q ^2-m^2}}\right)
-\mbox{Li}_2\left(
\frac{-2\sqrt{\Q ^2-m^2}}{\Q -\sqrt{\Q ^2-m^2}}\right) +
\frac{\pi^2}{6}\right.\nonumber\\ &-& \left. 2\pi
i\log\frac{4(\Q ^2-m^2)}{m^2} \right]\,,\nonumber\\ F(\Q ) &=&
2\left[\frac{\Q }{2\sqrt{\Q ^2-m^2}}
\left(\log\frac{\Q +\sqrt{\Q ^2-m^2}}{\Q -\sqrt{\Q ^2-m^2}} - 2\pi
i\right) -1\right] + 4\,. 
\eea 
We denoted here $N_\varepsilon^{IR}=\frac{1}{\epsilon}-\gamma_E +
\log(4\pi)$ and $k_0=v\cdot k = \frac12(k_++\frac{\vec k_\perp^2}{k_+})$. 
The IR singularity in
$\Delta_{IR}(\Q )$ originates from soft-gluon exchange between $b$
and $u$ in the initial state. The coefficient of
$N_\varepsilon^{IR}$ depends on the angle $\vartheta$ between the momenta
of the $b$ and $u$ quarks $\cosh\vartheta=(v\cdot k)/m$ and is well known 
as the QCD bremsstrahlung function. Notice that it receives an imaginary 
contribution due to the instantaneous (Coulomb) interaction.

The scale-dependent parameter $f_B(\mu_{\rm h})$ in (\ref{normal1}) 
is related to the physical decay constant $f_B$ by \cite{VoSh,PoWi} 
\beq\label{fBscaling}
f_B = \left(\frac{\alpha_s(m_b)}{\alpha_s(\mu_{\rm h})}
\right)^{-2/\beta_0} f_B(\mu_{\rm h})\,. \eeq
The logarithmic dependence on $\mu_h$ on the right-hand
side of (\ref{normal1}) matches that of the parameter $f_B(\mu_{\rm
h})$, as it should. The remaining mass-singular logarithm can be
absorbed into the wave function by introducing a factorization
scale $\Lambda^2$ satisfying $m \ll \Lambda_{QCD} \ll \Lambda
\ll \mu_{\rm h}$. Writing $\log(m^2/\mu_{\rm h}^2) =
\log(m^2/\Lambda^2) + \log(\Lambda^2/\mu_{\rm h}^2)$, the first
term is absorbed into the wave function and the second is resummed
into a factor similar to the one in (\ref{fBscaling})\footnote{
Note that this redefinition of the wave function applies strictly
for the purpose of the normalization condition. The precise 
$\Lambda$-dependence of the wave function $\tilde\psi(k_+,\Lambda)$ 
is derived below in Sec.~IV.}.

The IR singular terms can be resummed to all orders in
$\alpha_s$ using the QCD evolution equations. In the resulting
expression the contribution of multiple virtual soft gluon
emissions exponentiates and it can be factorized out from the wave
function.
This suggests to absorb the IR singular term $\Delta_{IR}$
(as well as the mass singular logarithm $\log(\Lambda^2/m^2)$ 
as explained above) into the light-cone wave function.
For the purpose of normalization alone, one can define thus a 
modified  wave function to one-loop order
\bea\label{psitilde}
\tilde \psi(k_+, k_\perp,\Lambda) &=&
\psi(k_+,\vec k_\perp)
\exp\left(
\frac{\alpha_s C_F}{4\pi}
\Delta_{IR}(v\cdot k)
+{\cal O}(\alpha_s^2)
\right)\left(1 + \frac{\alpha_s C_F}{4\pi}\cdot \frac32\log
\frac{\Lambda^2}{m^2}\right)
\eea
satisfying the normalization condition
\bea\label{normal3}
\sqrt{\frac{2}{N_c}} f_B m_B
\left(\frac{\alpha_s(m_b)}{\alpha_s(\Lambda)}\right)^{2/\beta_0} =
\int_0^\infty \mbox{d}k_+ \int\frac{d^2 \vec k_\perp}{(2\pi)^3}
\tilde \psi(k_+,k_\perp,\Lambda)
\left(1 + \frac{\alpha_s C_F}{4\pi}F(\Q )\right)\,.
\eea
It will be shown below that the hard scattering amplitude to one-loop order
is IR finite only when convoluted with this modified wave function.

\section{Leading twist analysis of $B\to \gamma\ell\nu_\ell$}

To leading order in $\alpha_s$ there are two diagrams contributing to the matrix
element (\ref{1}), shown in Fig.~1. Only the diagram (a), where the photon is
emitted from the light line, contributes to leading order in $1/m_b$.
Using the wave function (\ref{wf}), it can be written as
\bea\label{tree1}
\Gamma_\mu = Q_q \int\frac{dk_+ d^2 k_\perp}{2(2\pi)^3}
 \mbox{Tr}\left( \Psi(k_+,\vec k_\perp)
\gamma_\mu (1-\gamma_5)
\frac{\kslash - \pslash_\gamma}{-2k\cdot p_\gamma + i\varepsilon}
\epsslash^*\right)
\eea
The trace can be easily computed with the help of the basic relations
\bea
& &\mbox{Tr}\left\{ (k_+ + \vec\alpha_\perp\cdot \vec k_\perp)
\Lambda_+ P_v\gamma_5\,\, \gamma_\beta\gamma_5\right\}
= -v_+ (k_+ n_{+} + k_\perp)_\beta\,,\\
& &\mbox{Tr}\left\{ (k_+ + \vec\alpha_\perp\cdot \vec k_\perp)
\Lambda_+ P_v\gamma_5\,\, \gamma_\beta\right\}
= -\frac12 v_+ i\varepsilon(n_-,k_\perp,n_+,\beta)\,,
\eea
to which the expression (\ref{tree1}) can be reduced by application
of the identity
\beq
\gamma_\mu\gamma_\alpha\gamma_\nu =
g_{\mu\alpha} \gamma_\nu + g_{\nu\alpha} \gamma_\mu -
g_{\mu\nu} \gamma_\alpha + i\varepsilon(\mu,\nu,\alpha,\beta)
\gamma_\beta\gamma_5\,.
\eeq

We find in this way the following results for the form factors
to tree level
\bea\label{tree}
f_V(E_\gamma) = f_A(E_\gamma) = f_T(E_\gamma) =
Q_q
\sqrt{\frac{N_c}{2}} \frac{1}{E_\gamma}
\int_0^\infty
\frac{dk_+ d^2 k_\perp}{2(2\pi)^3} \frac{\psi(k_+,\vec k_\perp)}{k_+}
\left( 1 - \frac{\vec k^2_\perp}{2E_\gamma k_+}\right)\,.
\eea
The formfactor $f_T(E_\gamma)$ of the tensor current is encountered
when considering the radiative rare decay $B\to \nu\bar\nu \gamma$. It is defined
by the matrix element
\bea\label{T}
\frac{1}{\sqrt{4\pi\alpha}}
\langle\gamma(p_\gamma,\epsilon) |\bar b\sigma_{\mu\nu}\gamma_5 q| B(v)\rangle =
if_T(E_\gamma) [(p_\gamma)_\mu \epsilon^*_\nu - \epsilon_\mu^* (p_\gamma)_\nu ]\,.
\eea
(The tensor structure $\epsilon_\mu^* v_\nu - v_\mu \epsilon^*_\nu$ is forbidden by
gauge invariance.) The matrix element of the $\bar b\sigma_{\mu\nu} q$ current can
be obtained from this one with the help of the identity $\sigma_{\mu\nu}=
\frac{i}{2}\varepsilon_{\mu\nu\alpha\beta}\sigma^{\alpha\beta}\gamma_5$.

The corrections to the result (\ref{tree}) arising from the coupling
of the photon to the heavy quark (Fig.~1(b)) are suppressed by
$\Lambda/m_b$. In fact the leading term in this expansion is calculable
in terms of known quantities only. The corresponding correction is
given by
\bea
\Gamma^{(h)}_\mu = Q_b
\int\frac{dk_+ d^2 k_\perp}{2(2\pi)^3}
\mbox{Tr }\left(
\Psi(k_+,\vec k_\perp) \epsslash^* \frac{-m_b\vslash -\kslash'
+\pslash_\gamma + m_b}{(m_b v+k'-p_\gamma)^2-m_b^2}\gamma_\mu
(\gamma_5)\right)\,.
\eea
Performing the trace gives for the heavy quark contribution to the
form factors
\bea\label{masscorrs}
\delta^{(h)} f_i(E_\gamma) =
Q_b
\sqrt{\frac{N_c}{2}} \frac{1}{m_b E_\gamma}
\int_0^\infty
\frac{dk_+ d^2 k_\perp}{2(2\pi)^3} \psi(k_+,\vec k_\perp)\left(1 +
{\cal O}\left(\frac{\Lambda}{E_\gamma}, 
\frac{\Lambda^2}{m_b E_\gamma}\right)\right)
= Q_b \frac{f_B m_B}{2m_b E_\gamma}
\eea
where we used the normalization condition (\ref{normal}) for the wave
function. This correction is potentially important for the case
of charmed meson decays.

The equality of the form factors in (\ref{tree}) to lowest order
in $\alpha_s$ can be understood as the consequence of a larger
symmetry group of the Green functions in Fig.~1 to leading order
in $1/E_\gamma$.
To see this, one notes that the momentum of the light quark entering
the weak vertex
contains a large light-like component $p = k - p_\gamma =
-E_\gamma n_- + k$, with $n_\pm = (1,0,0,\pm 1)$.
Therefore a natural description of this
quark is in terms of the light-cone component of the quark field
$q_-$ defined as
\beq
q_- = e^{iE_\gamma (n_-\cdot x)} \Lambda_- q\,,\qquad
\Lambda_-=\frac{\gamma_+\gamma_-}{4}\,,
\eeq
satisfying $\Lambda_- q_- = q_-$ or $\nslash_- q= 0$.
The corresponding Dirac action reads, when expressed in terms
of this component \cite{DuGr}
\beq
\bar q(i\Dslash) q = \bar q_-(-in_-\cdot D) q_- + {\cal
O}(1/E_\gamma)\,,
\eeq
which contains an additional SU(2) symmetry group compared with
the original one.
This can also be seen in terms of the Feynman rules for the
light quark line
\bea
\mbox{propagator:}\qquad
& &i\frac{\kslash - \pslash_\gamma}{-2E_\gamma (n_-\cdot k) + k^2+
i\epsilon} = i\frac{-\nslash_-}{-2(n_-\cdot k)+i\epsilon} +
{\cal O}(1/E_\gamma)\\
\mbox{vertex:}\qquad
& &-ig \gamma_\mu t^a = -ig (-n_-)_\mu t^a + {\cal O}(1/E_\gamma)\,.
\eea

To leading order in $1/E_\gamma$ and $1/m_b$, the weak current
$\bar b\Gamma q$ can be written as $\bar h_v^{(b)}\Gamma q_-$, with
$h_v^{(b)}$ the static $b$ quark field satisfying $\gamma^0
h_v^{(b)}=-h_v^{(b)}$. Using the properties of the fields $q_-$ and
$h_v^{(b)}$ one can derive the following relation
\bea\label{Veff}
\bar h_v^{(b)}\gamma_\mu q_- &=& -(n_-)_\mu \bar h_v^{(b)} q_-
+ i\varepsilon_{0\alpha\mu\beta} (n_-)_\alpha
\bar h_v^{(b)} \gamma_\beta\gamma_5 q_- \,.
\eea
Taking the matrix element of (\ref{Veff}) between $\langle\gamma
(p_\gamma,\epsilon)|$ and $|B(v)\rangle$, and noting that
$\langle\gamma (p_\gamma,\epsilon)|\bar h_v^{(b)} q |B(v)\rangle=0$,
gives
\bea
\langle\gamma (E_\gamma n_-,\epsilon)|\bar h_v^{(b)}\gamma_\mu q_-
|B(v)\rangle = i\varepsilon_{0\alpha\mu\beta} (n_-)_\alpha
\langle\gamma (E_\gamma n_-,\epsilon)|\bar h_v^{(b)}\gamma_\beta
\gamma_5 q_-  |B(v)\rangle\,,
\eea
which reduces to $f_V(E_\gamma) = f_A(E_\gamma)$ in the rest frame
of $v$. Note that this is very different from other symmetry groups 
appearing in particle physics like flavor or spin as it is not apparent
in the hadron spectrum; rather it is a symmetry of an internal part
of a Feynman diagram mediating a decay process.
Similar arguments have been used in \cite{relations} to derive
relations among  semileptonic form factors in $B\to \pi,\rho$
using the additional symmetry of the so-called large
energy effective theory for the final state hadron \cite{DuGr}.

In the following we will show by explicit calculation to one-loop
order that the equality $f_V(E_\gamma) = f_A(E_\gamma)$ is preserved beyond tree
level, for the leading terms in an expansion of these form factors in powers of
$1/E_\gamma$.
Radiative corrections change the simple power law $1/E_\gamma$
by introducing a logarithmic dependence on the photon energy.
To leading order in $1/E_\gamma$ and $1/m_b$ these corrections are given by
(with $i=V,A,T$)
\bea\label{scaling}
f_i(E_\gamma) = \left(\frac{\alpha_s(m_b)}{\alpha_s(\mu_{\rm h})}\right)^{-2/\beta_0}
Q_q \sqrt{\frac{N_c}{2}} \frac{1}{2E_\gamma}
\int_0^\infty \mbox{d}k_+
\frac{\psi(k_+)}{k_+}\left(1 + \frac{\alpha_s C_F}{4\pi}
\ell_i(\mu_{\rm h}, E_\gamma, k_+)\right)
\eea
The first factor accounts for the different
renormalization of the weak current in the static quark
effective theory and QCD
\cite{VoSh,PoWi}.
The dependence on the hybrid renormalization scale $\mu_{\rm h}$ cancels between
this factor and the hard gluon correction in the effective theory $\ell_i$.

We consider in the following the one-loop radiative corrections to the diagram
in Fig. 1(a). The heavy-light vertex correction shown in Fig.~2(a) has the form
(for a general weak current $\bar b\Gamma q$)
\bea
\Lambda_\mu = -ig^2 C_F\left\{
[\Gamma]\left(\Sigma^{(a)} - (v\cdot\tilde q) J^{(a)}\right) -
[\Gamma (\tilde{\qslash} - \kslash )\vslash - \Gamma v\cdot(\tilde q-k)]
\langle (1-x)J^{(a)}\rangle \right\}\,.
\eea
The scalar integral $J^{(a)}$ is defined by (the definition of
$\langle (1-x)J^{(a)}\rangle$ is given in the Appendix)
\bea
J^{(a)}=\int\frac{\mbox{d}^4 l}{(2\pi)^4}
\frac{1}{(-v\cdot l+v\cdot k'+i\epsilon)
((l+\tilde q-k)^2+i\epsilon)(l^2+i\epsilon)}\,.
\eea
The heavy quark can be taken
on-shell such that its residual momentum $k$ satisfies $v\cdot k'=0$.
In fact the integrals
$J^{(a)}$ and $\langle (1-x)J^{(a)}\rangle$ are free of infrared and
collinear divergences.
The exact results for these integrals are presented in the Appendix in
(\ref{Jexact}), (\ref{(1-x)J}).
In the limit $k_+/E_\gamma \to 0$ they have the asymptotic expansions
\bea
J^{(a)} &=& \frac{i}{(4\pi)^2E_\gamma}\left\{
-\frac12 \log^2\frac{2E_\gamma}{k_+} -
\frac{2\pi^2}{3} + {\cal O}(\frac{k_+}{E_\gamma})\right\}\\
\langle (1-x)J^{(a)}\rangle &=& \frac{i}{(4\pi)^2E_\gamma}\left\{
-\frac12 \log^2\frac{2E_\gamma}{k_+} + \log\frac{2E_\gamma}{k_+}
-\frac{2\pi^2}{3} + {\cal O}(\frac{k_+}{E_\gamma})\right\}
\eea
The UV divergent integral $\Sigma^{(a)}$ is evaluated using
dimensional regularization in $D=4-2\varepsilon$ dimensions. One obtains
\bea
\Sigma^{(a)} = \frac{i}{(4\pi)^2}\left(
N_\varepsilon^{UV} - \log\frac{2E_\gamma k_+}{\mu^2_{\rm h}} + 2\right)
\eea
with $N_\varepsilon^{UV}=1/\varepsilon-\gamma_E+\log(4\pi)$.

Combining these results one finds the following contributions from the heavy-light
vertex correction to the $\ell_i$ factors from the diagram in Fig.~2(a)
\bea
\ell_V^{(a)} = \ell_A^{(a)} = N_\varepsilon^{UV}
-\log\frac{2E_\gamma k_+}{\mu^2_{\rm h}} -
\log^2 \frac{2E_\gamma}{k_+} + \log \frac{2E_\gamma}{k_+}
-\frac{4\pi^2}{3} + 2\,.
\eea

The light-light vertex correction shown in Fig.~2(b) introduces a correction
to the photon coupling of the form
\bea\label{light-light}
\epsslash^* &\to& \epsslash^* +
2ig^2 C_F \left\{ \epsslash^*
\left[ -(2-2\varepsilon) J^{(b)}_3 + 2(p_\gamma\cdot k)
(J^{(b)}-J^{(b)}_1+J^{(b)}_2 -J^{(b)}_5)\right]\right. \\
& & \qquad\qquad\qquad\qquad +\left. 2(\epsilon^*\cdot k)
\pslash_\gamma
\left[ -J^{(b)} + J^{(b)}_1 - J^{(b)}_2
+ J^{(b)}_5\right]\right\}\nonumber\,.
\eea
The scalar factors $J^{(b)}_i$ are defined by
\bea
(J^{(b)},J^{(b)}_{\mu}, J^{(b)}_{\mu\nu}) &=& \int \frac{\mbox{d}^4l}{(2\pi)^4}
\frac{(1,l_\mu, l_\mu l_\nu)}
{[(l+p_\gamma-k)^2 - m^2 + i\varepsilon]
[l^2-2l\cdot k+i\varepsilon](l^2+i\varepsilon)}\\
J^{(b)}_\mu &=& J^{(b)}_1 k_\mu + J^{(b)}_2 p_{\gamma\mu}\,,\\
J^{(b)}_{\mu\nu} &=& J^{(b)}_3 g_{\mu\nu} + J^{(b)}_4 k_\mu k_\nu +
 J^{(b)}_5(k_\mu  p_{\gamma\nu} +
 p_{\gamma\mu}k_\nu ) + J^{(b)}_6  p_{\gamma\mu} p_{\gamma\nu}\,.
\eea
These integrals have collinear singularities, which will be regulated by giving
the light quark a mass $m$. Their explicit results in the limit $m^2 \ll p_\gamma
\cdot k$ are given in the Appendix (see Eqs.~(\ref{J(b)})).
The term proportional to $(\epsilon^*\cdot k) \pslash_\gamma$ in the vertex correction
(\ref{light-light}) vanishes after the integration over
$\vec k_\perp$.
Keeping only the first term amounts to a multiplicative correction of the
lowest order result. Using the results Eq.~(\ref{J(b)}) one obtains the
following contributions to the $\ell_i$ coefficients from the diagram in
Fig.~2(b)
\bea\label{2b}
\ell_V^{(b)} = \ell_A^{(b)} = N_\varepsilon^{UV}
-\log\frac{2E_\gamma k_+}{\mu^2_{\rm h}}
+2 \log \frac{2E_\gamma k_+}{m^2} -1\,.
\eea
The self-energy correction on the internal light quark line (Fig.~2(c)) contributes
\bea
\ell_V^{(c)} = \ell_A^{(c)} = -N_\varepsilon^{UV} +\log\frac{2E_\gamma k_+}{\mu^2_{\rm h}}
- 1\,.
\eea
Finally, the box diagram (Fig.~2(d)) is given by
\bea\label{box}
B = ig^2 C_F\int\frac{\mbox{d}^4l}{(2\pi)^4}
\frac{\Gamma (\kslash+\lslash-\pslash_\gamma) \epsslash^*
(\kslash+\lslash)\vslash}
{(-v\cdot l+i\epsilon)((l+k-p_\gamma)^2-m^2+i\epsilon)
(l^2+2l\cdot k+i\epsilon) (l^2+i\epsilon)}\,.
\eea
The term of order $l^0$ in the loop momentum has an IR singularity,
which is regulated as before using dimensional regularization.
The total contribution of the box diagram to the $\ell$ coefficient
is given by (for both $i=V,A$)
\bea
\ell^{(d)} = -2i(4\pi)^2
\int\frac{d^Dl}{(2\pi)^D}
\frac{E_\gamma 2(v\cdot k) k_+ + E_\gamma (l_+ k_+ - k_\perp\cdot
l_\perp) + \frac12 l_-(k_\perp\cdot l_\perp - l_+ k_+)}
{(-v\cdot l+i\epsilon)((l+k-p_\gamma)^2-m^2+i\epsilon)
(l^2+2l\cdot k + i\epsilon)(l^2+ i\epsilon)}
\eea
Here the contribution of the $O(l^2)$ terms is of order $1/E_\gamma$
and thus subleading. The first two terms,
$\sim l^0$ and $\sim l^1$, can be computed to leading order in $E_\gamma$ by expanding the
large denominator as
$(l+k-p_\gamma)^2 - m^2\simeq -2E_\gamma (l_++k_+)$.
The numerator of the first two terms can be arranged as the sum of two
terms, one of which just cancels the denominator $l_++k_+$, plus
a remainder
\bea
E_\gamma 2(v\cdot k) k_+ + E_\gamma (l_+ k_+ - k_\perp\cdot
l_\perp) = E_\gamma \left( 2(v\cdot k)(l_++k_+)  +
(l_+ \frac{(k_\perp)^2}{k_+} - k_\perp\cdot l_\perp)
\right)\,.
\eea
The first term has exactly the structure of the scalar integral
appearing in the correction to $f_B$.
We obtain for the total contribution of the box diagram to leading order in $E_\gamma$
as
\bea\label{13}
\ell^{(d)} = i(4\pi)^2\left\{
2(v\cdot k) J_{IR}(v\cdot k) +
\int\frac{d^Dl}{(2\pi)^D}
\frac{l_+ \frac{(k_\perp)^2}{k_+} - k_\perp\cdot l_\perp}
{(-v\cdot l+i\epsilon)(l_++k_+)(l^2+2l\cdot k + i\epsilon)(l^2+ i\epsilon)}
\right\}
\eea
with $J_{IR}(\Q )$ the IR singular integral defined and computed in the
Appendix (see Eq.~(\ref{JIR(Q)})).
The second integral in (\ref{13}) is IR finite and can be easily
computed by combining the last two denominators with Feynman parameters.
This gives
\bea
& &\int\frac{d^4l}{(2\pi)^4}
\frac{(l_+, l_\perp)}
{(-v\cdot l+i\epsilon)(l_++k_+)(l^2+2l\cdot k + i\epsilon)(l^2+ i\epsilon)}
=\\
& &\int_0^1 dx
\int\frac{d^4l}{(2\pi)^4}
\frac{(l_+ - xk_+, l_\perp-xk_\perp)}
{(-v\cdot l+x\Q  + i\epsilon)(l_++k_+(1-x))(l^2 - m^2 x^2+ i\epsilon)^2}
=\nonumber\\
& &
-\frac{i}{8\pi^2}
\int_0^1 dx \int_0^\infty dl_+
\frac{(l_+ - xk_+, -xk_\perp)}
{[l_++k_+(1-x)][l_+^2 - 2xl_+ \Q  + m^2 x^2-i\epsilon]}\nonumber
\eea
where we performed the integration over the light-cone coordinates
$l_-, l_\perp$.
Inserting these results into the second integral in Eq.~(\ref{13})
one obtains
\bea
& &-\frac{i}{8\pi^2} \frac{(k_\perp)^2}{k_+}
\int_0^1 dx \int_0^\infty dl_+
\frac{l_+}
{[l_++k_+(1-x)][l_+^2 - 2xl_+ \Q  + m^2 x^2-i\epsilon]}\\
&=& -\frac{i}{(4\pi)^2}\left(\log^2\frac{k_+}{2\Q } +2\pi i
\log\frac{k_+}{2\Q }\right)\,.\nonumber
\eea
Finally, we obtain the following result for $\ell^{(d)}$ to leading twist
\bea\label{boxtotal}
\ell_V^{(d)} = \ell_A^{(d)} = i(4\pi)^2\, 2\Q J_{IR}(\Q ) +
\log^2\frac{k_+}{2\Q } + 2\pi i\log\frac{k_+}{2\Q } \,.
\eea

\begin{center}
\begin{tabular}{|c|l|}
\hline
\hline
Diagram   &  Contributions to $\ell_i(E_\gamma)$ \\
\hline
2(a) & $N_\varepsilon^{UV} - \log\frac{2E_\gamma k_+}{\mu^2}
  - \log^2\frac{2E_\gamma}{k_+} + \log\frac{2E_\gamma}{k_+} -
  \frac{4\pi^2}{3} + 2$\\
2(b) & $N_\varepsilon^{UV} - \log\frac{2E_\gamma k_+}{\mu^2} +
  2\log\frac{2E_\gamma k_+}{m^2} -1$ \\
2(c) & $-N_\varepsilon^{UV} + \log\frac{2E_\gamma k_+}{\mu^2} - 1$\\
2(d) & $i(4\pi)^2\, 2\Q J_{IR}(\Q ) +
\log^2\frac{k_+}{2\Q } + 2\pi i\log\frac{k_+}{2\Q }$ \\
$\frac12(Z_2^{QCD}-1)$ & $-\frac12 N_\varepsilon^{UV} +
\frac32\log\frac{m^2}{\mu^2} - N_\varepsilon^{IR}$ \\
$\frac12(Z_2^{HQET}-1)$ &  $N_\varepsilon^{UV}-N_\varepsilon^{IR}$ \\
\hline
Total & $\frac32 N_\varepsilon^{UV} -
\log\frac{2E_\gamma k_+}{\mu^2} +
\frac12\log\frac{\mu^2}{m^2}$ \\
 & $- \log^2\frac{2E_\gamma}{k_+} + \log\frac{2E_\gamma}{k_+} +
2\log\frac{2E_\gamma k_+}{m^2} - \frac{4\pi^2}{3}$ \\
 & $+ \Delta_{IR}(\Q )
 +\log^2\frac{k_+}{2\Q } + 2\pi i\log\frac{k_+}{2\Q }$ \\
\hline
\hline
\end{tabular}
\end{center}
\begin{quote} {\bf Table 1.}
One-loop contributions to the form factor from individual diagrams.
The IR singular contribution $\Delta_{IR}(\Q ) = i(4\pi)^2
2k_0 J_{IR}(k_0) - 2(N_\varepsilon^{IR}+\log\frac{\mu^2}{m^2})$ is 
identical to the one appearing in the one-loop correction to $f_B$ 
(\ref{JIR}) and can be absorbed into the $B$ meson light-cone wave 
function as explained in Sec.~II.
\end{quote}

We are now in a position to write down the complete one-loop correction
to the form factors for $B\to \ell\bar\nu_\ell \gamma$.
The individual contributions from the diagrams of Fig.~2 and their
total result are presented in Table 1. There are a few remarks which
can be made about these results.

\begin{itemize}
\item The box diagram (Fig.~2(d)) contains an IR divergent term
(\ref{boxtotal}) which depends on $\vec k_\perp$ through the
quantity $v\cdot k$. Note that this is different from the case
of the pion form factor, which is IR finite \cite{Braaten}, and
contains only mass singularities.
However, the IR singular term can be seen to be precisely identical to the
one appearing in the one-loop correction Eq.~(\ref{normal1}) to the
decay constant $f_B$. As explained in Sec.~II, it can be absorbed
into the $B$ meson light-cone wave function, leaving
a IR-finite Wilson coefficient depending only on the
light-cone momentum component $k_{+}$.

\item The dependence on the $\overline{MS}$ hybrid scale
$\mu_{\rm h}$ cancels,
as it should, between the $\ell$ coefficient and the
corresponding factor in (\ref{scaling}). We will choose for
this scale $\mu=2E_\gamma$, with which the first factor accounts
explicitly for the large logs $(\frac{\alpha_s}{\pi}
\log(\frac{m_b}{2E_\gamma}))^n$ in
leading logarithmic approximation.

\item
The equality of the leading twist form factors for different
currents noted at tree level $f_V(E_\gamma)= f_A(E_\gamma)$
persists to one-loop order. In view of the symmetry arguments
justifying this equality at tree level, it is tempting to conjecture 
that this is a general result for the leading twist form factors,
valid to all orders in the strong coupling.
\end{itemize}

With these remarks, the leading twist result for the form factors
in $B\to \ell\bar \nu_\ell\gamma$ decays can be written as\footnote{
The modified wave function $\tilde \psi(k_+,k_\perp)$ in this
expression contains only the exponentiated IR singularity.} 
\bea\label{scalingV}
& &f_{V,A}(E_\gamma) =
Q_q \sqrt{\frac{N_c}{2}}
\left(\frac{\alpha_s(m_b)}{\alpha_s(2E_\gamma)}
\right)^{-2/\beta_0}
\frac{1}{E_\gamma}
\int_0^\infty \frac{\mbox{d}k_+  \mbox{d}^2\vec k_\perp}
{2(2\pi)^3}
\tilde \psi(k_+,\vec k_\perp) 
T_H(k_+,\vec k_\perp)
\eea
where the hard scattering kernel $T_H(k_+,\vec k_\perp)$ is 
given to one-loop order by
\bea\nonumber
T_H(k_+,\vec k_\perp) &=&
\frac{1}{k_+}
\left(1 + \frac{\alpha_s(2E_\gamma) C_F}{4\pi}
\left\{
- \log^2\frac{2E_\gamma}{k_+} + \frac52\log\frac{2E_\gamma}{k_+}
+ \frac52 \log\frac{2E_\gamma k_+}{m^2} 
\right.\right.\\\label{TH}
& &\left.\left.
- \frac{4\pi^2}{3}
+\log^2\left(1 + \frac{\vec k_\perp^2}{k_+^2}\right)
 - 2\pi i\log\left(1 + \frac{\vec k_\perp^2}{k_+^2}\right)\right\}
\right) \,.
\eea
Note that this result for the form factors is sensitive to the
dependence of the wave function on the transverse momenta,
through the last two terms. After integration over $\vec k_\perp$,
these terms will give a finite correction to the light-cone
wave function. The last term in (\ref{TH}) will give the
form factors also a complex phase.
These features are in contrast to the pion form factor case,
where transverse momentum effects are absent to leading twist.

\section{Mass-singular logarithms and Sudakov effects}

The expression for the form factors (\ref{scalingV}) contains
mass-singular logarithms $\log(2E_\gamma k_+/m^2)$ as well as 
Sudakov double logs 
$\log^2(2E_\gamma/k_+)$ which must be resummed to all orders. 
In this section we will discuss these issues in turn.

\subsection{Resummation of collinear singularities}

The hard scattering amplitude $T_H(k_+)$
contains collinear logs $\log(2E_\gamma k_+/m^2)$, which
arise only from the diagram 2(b) and the wave function
renormalization constant. In the former, these logs are produced by
integration over the transverse momenta in the region
 $m^2 \le \vec l_\perp^2 \le 2E_\gamma k_+$. The propagator of 
the struck quark can be written in this region as
$(k-p_\gamma+l)^2 \simeq -2E_\gamma (k_++l_+)+\cdots$. 
Keeping only the leading terms in $1/E_\gamma$, the contribution of 
this diagram (plus the tree contribution) is proportional to
\bea\label{masssingraw}
& &\Gamma^{(0)}+\Gamma^{(b)} = \\
& &\int dk_+ \psi(k_+) \left(\frac{1}{k_+} +
\frac{\alpha_s C_F}{4\pi}\left(
\int dl_+\frac{2}{(k_+)^2}\theta(k_+-l_+) +
\frac{1}{2k_+}\right)\log\frac{2E_\gamma k_+}{m^2}\right) =\nonumber\\
& &\int dk_+dl_+ \psi(k_+) \left(\delta(k_+-l_+) + {\cal K}(k_+,l_+)
\int_{m^2}^{2E_\gamma k_+}\frac{d\vec l_\perp^2}{\vec l_\perp^2}
\right) T_H^{(0)}(l_+) \,,\nonumber
\eea
where $T_H^{(0)}(l_+)=1/l_+$ is the tree-level hard scattering 
amplitude. The kernel ${\cal K}(k_+,l_+)$ is given by
\bea
{\cal K}(k_+,l_+) =
 \frac{\alpha_s C_F}{4\pi}\left[\left(
\frac{2l_+\theta(k_+-l_+)}{k_+(k_+-l_+)}\right)_+
+ \frac12 \delta(k_+-l_+) \right]
\,,
\eea
where the term proportional to $\delta(k_+-l_+)$ comes from the wave
function renormalization.
The $+$-distribution is defined as usual by
\beq
f(k_+,l_+)_+ = f(k_+,l_+) - \delta(k_+-l_+)\int dr_+ f(k_+,r_+)\,.
\eeq

Integrating over $l_+$ gives the explicit one-loop result for the
mass-singular logs (\ref{2b}). However, writing the result in this 
form helps us to resum these logs to all orders. To do this, one cuts 
the integral over the transverse loop momentum in (\ref{masssingraw})
to a certain cut-off $\Lambda$. This will be chosen identical to the
one introduced in the normalization condition (\ref{normal3}). 
The logarithm resulting from integration over the range 
$m^2 < \vec l_\perp^2 < \Lambda^2$ is then absorbed into the
wave function $\tilde \psi(k_+,\Lambda)$ by defining
\bea\label{shuffle}
\tilde \psi(l_+,\Lambda^2) = \int dk_+ 
\psi(k_+) \left(\delta(k_+-l_+) +
{\cal K}(k_+,l_+)
\log\frac{\Lambda^2}{m^2}\right)\,.
\eea

Expressed in terms of the  wave function $\tilde \psi(k_+,\Lambda)$, 
the form factor is written as
\bea\label{ffmsing}
& &f_{V,A}(E_\gamma) =
Q_q \sqrt{\frac{N_c}{2}}
\left(\frac{\alpha_s(m_b)}{\alpha_s(2E_\gamma)}
\right)^{-2/\beta_0}
\frac{1}{E_\gamma}
\int_0^\infty \mbox{d}k_+
\frac{\tilde \psi(k_+, 2E_\gamma k_+)}{k_+}\\
& & \times\left(1 + \frac{\alpha_s(2E_\gamma) C_F}{4\pi}
\left\{
- \log^2\frac{2E_\gamma}{k_+} + \frac52\log\frac{2E_\gamma}{k_+}
 - \frac{4\pi^2}{3}
+\log^2\frac{k_+}{2\Q} + 2\pi i\log\frac{k_+}{2\Q}\right\} \right)
\,.\nonumber
\eea

Taking the logarithmic derivative of (\ref{shuffle}) with respect
to $\Lambda^2$ gives the integral equation satisfied by $\tilde 
\psi(l_+,\Lambda^2)$
\bea
\Lambda^2\frac{d}{d\Lambda^2} \tilde \psi(l_+,\Lambda^2) =
\int_0^\infty dk_+ {\cal K}(k_+,l_+)
\tilde \psi(k_+,\Lambda^2)\,.
\eea
This is the analog of the Brodsky-Lepage evolution equation which
governs the evolution of 
the light-cone wave function of a heavy meson with the factorization
scale. The moments of the wave function are renormalized 
multiplicatively with the anomalous dimensions
\bea\label{moments}
\Lambda^2\frac{d}{d\Lambda^2} \langle k_+^n\rangle =
\frac{\alpha_s C_F}{4\pi}
\left(\frac{2}{n+2}+\frac12\right) \langle k_+^n\rangle\,,
\eea
where $\langle k_+^n\rangle = \int dk_+
(k_+)^n\tilde\psi(k_+,\Lambda)$. The 0$^{\rm th}$ moment of the 
wave function 
evolves with the same anomalous dimension $3\alpha_s C_F/(8\pi)$
as previously derived in Sec.~II (see Eq.~(\ref{normal3})).

\subsection{Sudakov resummation}

The radiative correction to the form factors $f_i(E_\gamma)$
contains double logarithms of the large ratio $\log^2\frac{2E_\gamma}
{k_+}$. The explicit calculation of the preceding section shows that
such logarithms arise (in the Feynman gauge) from one loop correction
to the vertex $b\to W q$ of the weak decay of the $b-$quark into a light
quark with momentum $Q=p_\gamma-k$. It is easy to see that in the
rest frame of the $B-$meson in the kinematical region $E_\gamma \gg
\Lambda_{QCD}$
the light quark moves close to the ``+'' light-cone direction along the
photon momentum with the energy $(Q\cdot v)\sim E_\gamma$ and small virtuality
$Q^2\sim -2(p_\gamma\cdot k)$ such that
${Q^2}/{(2Q\cdot v)^2} = {\cal
O}\left({\bar\Lambda}/{E_\gamma}\right)$.
It is the ratio of the scales ${Q^2}/{(2Q\cdot v)^2}\ll 1$
that enters as an argument into Sudakov double logs. The
appearance of large negative corrections is related to
enhancement of the contribution of soft virtual gluons
propagating collinear to the produced light quark, close to the direction
of photon momentum. In contrast with the inclusive distributions where
it is canceled against the contribution of real soft gluon emissions,
virtual soft gluon contribution survives for an exclusive 
distribution like the one under consideration due to the absence of 
real soft gluons in the final states.

Let us consider the one-loop Sudakov correction to the weak decay vertex
\begin{equation}
F=1-\frac{\alpha_s}{4\pi} C_F S\,,\qquad
S=\log^2\frac{4(Q\cdot v)^2}{-Q^2}-
\log\frac{4(Q\cdot v)^2}{-Q^2}
\end{equation}
with $2(Q\cdot v)=2E_\gamma$ and $Q^2=-2E_\gamma k_+$ in the rest frame of the B-meson.
The Sudakov form factor $S$ is given by the following one-loop
Feynman integral
\begin{equation}
S=i\int \frac{d^4l}{(2\pi)^2}
\frac{(4Q_--2l_-)v_+}{(-v\cdot l+i\epsilon)(l^2+i\epsilon)((Q-l)^2+i\epsilon)}
\end{equation}
with $Q_-=2E_\gamma$, $Q_+=-k_+$, $Q^2=Q_+Q_-$ and $v_+=1$.
Calculating $S$ we write the integration measure as
$d^4l=\frac12dl_+dl_-d^2l_\perp$ and perform $l_+-$integration by taking
residues at the poles corresponding to three different
propagators. One finds that the integral is different from zero
provided that $l_-$ belongs to one of the regions, $0< l_- < Q_-$
and $l_- > Q_-$. One checks that in the second case the gluon has
large components of the momenta and its contribution is associated
with short distance (hard) subprocess. In the first case, the
$l_+-$integral is given by the residue at $l_+=l_\perp^2/(2l_-)
-i\epsilon$ which effectively amounts to putting the virtual gluon
on-shell, $l^2=0$. Then, introducing the scaling variable
$x=l_-/Q_-$ one finds
\begin{equation}
S=\frac12\int_0^1\frac{dx}{x}\int_0^\infty dl_\perp^2
\frac{4-2x}{\left(x+\frac{l_\perp^2}{x(2E_\gamma)^2}\right)
\left(2E_\gamma k_+(1-x)+\frac{l_\perp^2}{x}\right)}\,.
\end{equation}
The denominators effectively set the limits on the integration
ranges, such that the leading doubly logarithmic correction arises
from the region (in the rest frame of the $B$ meson)
\beq
k_+ \leq l_- \leq 2E_\gamma \,,\qquad
k_+ l_- \leq l_\perp^2 \leq l_-^2\,.
\eeq
In this way, one calculates the one-loop correction to the Sudakov
form factor as
\begin{eqnarray}
S
&=&\int_{k_+}^{2E_\gamma} \frac{dl_-}{l_-} \int_{k_+l_-}^{l_-^2}
\frac{dl_\perp^2}{l_\perp^2}
\left(2-\frac{l_-}{2E_\gamma}\right)
\nonumber
\\
&=&\int_{k_+^2}^{2k_+E_\gamma}\frac{dl_\perp^2}{l_\perp^2}\ln\frac{l_\perp^2}{k_+^2}
+\int^{(2E_\gamma)^2}_{2k_+E_\gamma}\frac{dl_\perp^2}{l_\perp^2}\ln\frac{(2E_\gamma)^2}{l_\perp^2}
-\int^{(2E_\gamma)^2}_{2E_\gamma k_+}\frac{dl_\perp^2}{l_\perp^2}
\label{1-loop}
\end{eqnarray}
The reason why we represented the one-loop correction in this particular
form is that it admits generalization to higher orders in the coupling
constant that effectively resums Sudakov logarithms.
Each term in the r.h.s.\ of (\ref{1-loop}) comes from different part of gluon phase
space and has the following interpretation. The last term
describes collinear emission of on-shell energetic
gluon, $l_-={\cal O}(E_\gamma)$, $l_+\ll l_-$ and $l_\perp^2=l_+l_-={\cal
O}(k_+E_\gamma)$ (collinear region), and gives rise to a single collinear log. The first two
terms correspond to soft gluon emission on two different infrared
scales, $l_+\sim l_- \sim l_\perp = {\cal O}(k_+)$ (soft region) and
$l_+\sim l_- \sim l_\perp = {\cal O}(\sqrt{k_+E_\gamma})$ (infrared region),
and produce double logarithmic contributions. Examining higher order
corrections to the $b\to W q$ vertex one can show that the
same regions of gluon momenta provide the dominant contribution to the
Sudakov form factors. Moreover, since the emission of collinear
and soft gluons occurs on different time scale their contribution
factorizes out as \cite{GKPLB}
\beq
F=F_H(Q_+^2,\mu^2) F_J(Q_+^2,Q_+Q_-,\mu^2) F_S(Q_+^2,Q_+Q_-,\mu^2) F_{IR}(Q_+Q_-,Q_-^2,\mu^2)
\label{fact}
\eeq
with $Q_-=2E_\gamma$ and $Q_+=-k_+$. Here the hard subprocess $F_H$
takes into account short distance
corrections to the weak decay vertex, $l_\mu \sim Q_+$, while
$F_J$, $F_S$ and $F_{IR}$ denote contributions of collinear and
soft gluons on different momentum scales. The parameter $\mu$
entering (\ref{fact}) plays the role of the factorization scale.
The subprocesses $F_S$ and $F_{IR}$ admit an operator definition
as expectation values of the Wilson lines originating as eikonal
phase of quarks interacting with soft gluons. Using this
interpretation one can show that the $\mu-$evolution of $F_S$ and
$F_{IR}$ subprocesses is in one-to-one correspondence with the
renormalization properties of Wilson lines. In this way, using
evolution equations for different subprocesses and
the $\mu-$independence of $F$, one finds that the Sudakov form factor 
obeys the following evolution equation
\beq
\frac{d \ln F}{d\ln E_\gamma}=\Gamma(\alpha_s(Q_+^2)) +
\Gamma_0(\alpha_s(Q_-^2)) - \frac12 \int_{2E_\gamma k_+}^{(2E_\gamma)^2}
\frac{dl_\perp^2}{l_\perp^2} \Gamma_{\rm
cusp}(\alpha_s(l_\perp^2))\,.
\label{EQ}
\eeq
This evolution equation involves three functions of the coupling constant
that appear as anomalous dimensions in the evolution equations for different
subprocesses. Two of them $\Gamma_{\rm cusp}$ and $\Gamma_0$ are related
to renormalization of (light-like) Wilson loops while $\Gamma$ is
related to the UV renormalization of the weak decay vertex
\beq
\Gamma_{\rm cusp}= \frac{\alpha_s}{\pi}C_F+ {\cal O}(\alpha_s^2)
\,,\quad
\Gamma_0 = 0 + {\cal O}(\alpha_s^2)
\,,\quad
\Gamma=\frac{\alpha_s}{\pi}C_F+ {\cal O}(\alpha_s^2)\,.
\label{Gamma's}
\eeq
Neglecting the $\Gamma_0-$term one can write the solution to the
evolution equation (\ref{EQ}) as
\beq
-4\ln F=\int_{Q_-^2}^{Q^2}\frac{dl_\perp^2}{l_\perp^2}\ln\frac{l_\perp^2}{Q_-^2}\Gamma_{\rm
cusp}(\alpha_s(l_\perp^2))
+\int^{Q_+^2}_{Q^2}\frac{dl_\perp^2}{l_\perp^2}\ln\frac{Q_+^2}{l_\perp^2}\Gamma_{\rm
cusp}(\alpha_s(l_\perp^2))
-\int^{Q_+^2}_{Q^2}\frac{dl_\perp^2}{l_\perp^2}\Gamma(\alpha_s(l_\perp^2))
\label{sol-EQ}
\eeq
Comparing (\ref{sol-EQ}) with the one-loop expression (\ref{1-loop}) we conclude
that, first, Sudakov logarithms exponentiate \cite{CoSo} and, second, the
exponent of the Sudakov form factor is formally given by the one-loop
expression in which the ``bare'' QCD coupling constant is replaced by
an anomalous dimension with a particular choice of the normalization
scale given by gluon transverse momentum $l_\perp^2$. The perturbative
expansion (\ref{sol-EQ}) is valid provided that the integration over
$l_\perp^2$ does not go below the Landau singularities of the coupling
constant. This means that the resummed expression (\ref{sol-EQ}) is
valid provided that $k_+^2 \gg \Lambda_{\rm QCD}$. In the practical
application discussed below, the Sudakov form factor for $k_+$ below
the singularity at $k_+=\Lambda_{\rm QCD}$ will be frozen at its value
just above this point.

Expanding the QCD running coupling constant $\alpha_s(l_\perp^2)$
in powers of $\alpha_s(E_\gamma)$ and performing the integration in
(\ref{sol-EQ}) one can expand the exponent $\ln F$ into a series of
the form $\alpha_s(\alpha_s L^2)^n$, $\alpha_s^2(\alpha_s L^2)^n$,
$\dots$ with $n=1\,,2\,,\dots$, $\alpha_s=\alpha_s(E_\gamma)$ and $L=\ln(E_\gamma/k_+)$
to which we shall refer as leading-order (LO), 
next-to-leading order(NLO), $\dots$ corrections. In particular, to 
the LO approximation it proves enough to keep only the first two 
terms in (\ref{sol-EQ}). Using the one-loop running of the strong 
coupling
\bea
\alpha_s(\vec l_\perp^2)
=\frac{4\pi}{\beta_0\ln\frac{l_\perp^2}{\Lambda^2}}
\eea
with $\beta_0=11-\frac23 n_f$ and replacing $\Gamma_{\rm cusp}$ by its one-loop
expression, one gets
\beq
S_{\rm LO} = \frac{2C_F}{\beta_0}
\left\{
-\log\frac{2E_\gamma k_+}{\Lambda^2}
\log\left[\frac12\log\frac{2E_\gamma k_+}{\Lambda^2}\right]
+ \log\frac{2E_\gamma}{\Lambda}
\log\log\frac{2E_\gamma}{\Lambda} +
\log\frac{k_+}{\Lambda} \log\log\frac{k_+}{\Lambda}
\right\}\,.
\eeq
which agrees with the result in \cite{AkStYa}.
The overall effect of the Sudakov form factor is to depress
the form factors at large values of $E_\gamma$.
One can systematically improve the accuracy of (\ref{sol-EQ}) by taking
into account NLO terms. To this end one should include two-loop
corrections to the coupling constant and $\Gamma_{\rm cusp}$
as well as one-loop correction to the anomalous dimensions
$\Gamma$ defined in (\ref{Gamma's}).

\section{Application}

The decay rate for $B\to \gamma \nu_\ell\ell^+$ differential
in the lepton and photon energy is
\bea
\frac{\mbox{d}^2\Gamma}{\mbox{d}E_e \mbox{d}E_\gamma} &=&
\frac{\alpha G_F^2 |V_{ub}|^2 m_B^3}{4(2\pi)^2}
\left\{ [f_A^2(E_\gamma)+f_V^2(E_\gamma)](-2xy+2xy^2+x-2x^2 y+x^3)\right.\\
& &\left.\qquad\qquad
-2f_A(E_\gamma) f_V(E_\gamma) x(1-x)(1+x-2y)\right\}\nonumber\,.
\eea
We denoted here $x=1-2E_\gamma/m_B$ and $y=2E_e/m_B$, in terms of
which the available
phase space is described as $x=(0,1)$ and $y=(x,1)$. An integration
over all possible
values of the electron energy $y$ gives for the rate as function of
the photon energy
\bea
\frac{\mbox{d}\Gamma}{\mbox{d}E_\gamma} &=&
\frac{\alpha G_F^2 |V_{ub}|^2 m_B^4}{12(2\pi)^2}
[f_A^2(E_\gamma)+f_V^2(E_\gamma)] x(1-x)^3\,.
\eea
Our results for the form factors $f_{V,A}(E_\gamma)$ can be
therefore turned into a prediction for the shape of the photon
spectrum in this decay. To leading twist, the $1/E_\gamma$
dependence of these form factors yields a symmetrical
photon spectrum $d\Gamma/dx \propto x(1-x)$.

Neglecting radiative corrections, the form factors parametrizing
the $B^+\to \ell^+ \nu \gamma$ decay are given by
\bea\label{qmff}
f_{V,A}(E_\gamma) =
\frac{f_B m_B}{2E_\gamma} (Q_u R - \frac{Q_b}{m_b}) +
{\cal O}\left(\frac{\Lambda^2}{E_\gamma^2}\right)
\,,
\qquad R\equiv \frac{\langle (k_+)^{-1}\rangle}
{\langle (k_+)^0\rangle}\,,
\eea
where we included also the leading $\Lambda/m_b$ correction computed
in (\ref{masscorrs}).

Extrapolating the tree-level form factors (\ref{qmff}) over
the entire phase space gives for the integrated decay rate
\bea\label{inclusive}
\Gamma (B^+\to \ell\nu\gamma) =
\alpha \frac{G_F^2 |V_{ub}|^2m_B^5}{288\pi^2}
f_B^2 \left(Q_u R - \frac{Q_b}{m_b}\right)^2\,.
\eea
This result is identical to the one obtained in \cite{AES} from a
quark model calculation of the annihilation graph, with the
identification $R\to 1/m_u$ (the inverse constituent quark mass).
In fact the appearance of the inverse constituent quark mass is a
common aspect of quark model calculations of long distance effects
produced by weak annihilation topologies with emission of
one photon or gluon \cite{BSS}. Such contributions have been 
investigated in many processes such as $B\to 
\rho\gamma$ \cite{AtBlSo} and $B\to D^*\gamma$ \cite{Cheng}.

Our QCD-based derivation gives such computations a precise meaning
by replacing the ambiguous notion of constituent quark mass
with a well-defined integral over the light-cone $B$ meson wave
function. Besides specifying the limits of validity of this result,
such an approach allows one to compute also strong interactions
corrections to it in a systematic way.

It is possible to derive a model-independent lower limit on the 
magnitude of the $R$ parameter, under the assumption that the
light-cone wave function is everywhere positive, which is reasonable
for the ground state $B$ meson. This bound reads
\beq\label{bound}
R \geq \frac{\langle k_+^0\rangle}{\langle k_+\rangle} =
\frac{3}{4\bar\Lambda}\,,
\eeq
and can be proved with the help of the inequality
\beq
\frac{1}{k_+} + b k_+ \geq 2\sqrt{b}\,,\qquad k_+ > 0\,.
\eeq
Here $b$ is an arbitrary real positive number. Multiplying with
$\psi(k_+)$ and integrating over $k_+$ gives the inequality 
$R \geq 2\sqrt{b} - \frac43\bar\Lambda b$, where we used the
normalization condition (\ref{n=1}). This is most restrictive provided
that one chooses $\sqrt{b} = \frac{3}{4\bar\Lambda}$, which gives the
result (\ref{bound}).

It is interesting to note that a hadronic parameter related
to $R$ appears also in the description of the nonfactorizable corrections
to nonleptonic $B\to\pi\pi$ decays \cite{BBNS} (called there $1/\lambda_B$).
Our results suggest therefore a method for extracting this
parameter in a model-independent way from data on
$B\to \gamma e\nu$ decays.

To eliminate the dependence on $f_B$ and $V_{ub}$, we will present
our results for the photon spectrum by normalizing it to the
pure leptonic decay rate for $B^+\to \mu\nu$, which is given by
\bea
\Gamma_l(B^+\to \mu\nu) =
\frac{G_F^2 |V_{ub}|^2 m_B^3}{8\pi}
f_B^2\left(\frac{m_\mu}{m_B}\right)^2
\left(1 - \frac{m_\mu^2}{m_B^2}\right)\,.
\eea

For illustrative purposes we will adopt in the following numerical
estimates a two-parameter Ansatz for the heavy meson
light-cone wave function inspired by the oscillator model of \cite{BSW}
\bea
\psi(k_+) = {\cal N} k_+ \exp\left(
-\frac{1}{2\omega^2}(k_+ - a)^2\right)\,.
\eea
We will vary the width parameter $\omega$ in the range 
$\omega = 0.1-0.3$ GeV. The parameters ${\cal N}$ and $a$ will
be determined from the normalization conditions discussed in Sec.~II.
For a given value of $\bar\Lambda$, these normalization conditions set
an upper bound on the width parameter $\omega$, given by 
$\omega_{max}=\frac{8}{3\sqrt{2\pi}} \bar\Lambda$ 
(corresponding to $a=0$). The latter will be taken 
between $\bar\Lambda=0.3$ GeV and 0.4 GeV.
The resulting numerical value of the constant $R$ together with the parameter
$a$ are given in Table 2 for several choices of $\bar\Lambda$
and $\omega$.

\begin{center}
\begin{tabular}{|c|c|c|c|c|c|c|c|c|c|}
\hline
  & \multicolumn{3}{|c|}{$\bar\Lambda=0.3$ GeV}
  & \multicolumn{3}{|c|}{$\bar\Lambda=0.35$ GeV}
  & \multicolumn{3}{|c|}{$\bar\Lambda=0.4$ GeV} \\
\cline{2-10}
$\omega$ (GeV) & 0.1 & 0.2 & 0.3 & 0.1 & 0.2 & 0.3 & 0.1 & 0.2 & 0.3 \\
\hline\hline
$a$ (GeV)  & 0.37 & 0.27 & 0.05 & 0.44 & 0.36 & 0.19 & 0.51 & 0.44 & 0.30 \\
\hline
$R$ (GeV$^{-1}$) & 2.70 & 3.29 & 3.87 & 2.28 & 2.65 & 3.09 & 1.96 & 2.23
 & 2.59 \\
\hline
\end{tabular}
\end{center}
\begin{quote} {\bf Table 2.} Light-cone wave function parameters
$a$ and $R$ corresponding to several values of the
binding energy $\bar\Lambda$ and the width parameter $\omega$.
\end{quote}

Taking $f_B=175$ MeV and $|V_{ub}|=3.25\times 10^{-3}$ \cite{Vub,Lig}
gives for the muonic $B^+$ decay mode a branching ratio
\bea
{\cal B}(B^+\to \mu\nu) = 2.3\times 10^{-7}\,.
\eea
For a typical range of values $R=2-3$ GeV$^{-1}$ (see Table 2),
the tree-level
integrated rate (\ref{inclusive}) predicts a ratio
\bea\label{treeratio}
\frac{{\cal B}(B^+\to \gamma e^+\nu)}{{\cal B}(B^+\to \mu\nu)}
=2.02 R^2 \simeq 8 - 18\,,
\eea
which implies branching ratios of about $(2-5)\times 10^{-6}$ for
the $B^+$ radiative leptonic mode, in agreement with the general
estimates of \cite{AES}.

A similar analysis can be made for the radiative leptonic decay
$D^+\to \gamma e^+\bar\nu$, for which one obtains the ratio of
branching ratios
\bea\label{treeratioD+}
\frac{{\cal B}(D^+\to \gamma e^+\nu)}{{\cal B}(D^+\to \mu\nu)}
= 0.07 \left(Q_d R-\frac{Q_c}{m_c}\right)^2
\simeq 0.09 - 0.16\,.
\eea
Note that the charmed
quark contribution can be appreciable, and can account for up to
50\% of the light quark contribution.
Neglecting SU(3)  breaking effects and small kinematical corrections,
the denominator can be related to the muonic branching
ratio for $D_s^+$ decay which has been measured
\beq
{\cal B}(D^+\to \mu^+\nu) = \left(\frac{V_{cd}}{V_{cs}}\right)^2
\frac{\tau(D^+)}{\tau(D_s^+)} {\cal B}(D_s^+\to \mu^+\nu) \simeq
(0.68\pm 0.37)\times 10^{-3}\,.
\eeq
We used here the CLEO result 
${\cal B}(D_s\to \mu^+\nu) = (6.2\pm 3.1)\times 10^{-3}$ \cite{CLEO}.
This predicts an absolute branching ratio for the radiative $D^+$
decay of
\beq
{\cal B}(D^+\to \gamma e^+\nu) = (0.82\pm 0.65)\times 10^{-4}\,.
\eeq
Somewhat larger absolute values are obtained for the $D_s^+$
radiative decay width, which is enhanced by the larger CKM matrix element
$V_{cs}$. Neglecting SU(3) breaking in the hadronic parameter
$R$ one finds for this case
${\cal B}(D_s^+\to \gamma e^+\nu) = (0.9\pm 0.8)\times 10^{-3}$,
again in agreement with the estimates of \cite{AES}.


While useful as an order of magnitude estimate, we stress that
the relation (\ref{inclusive}) and the numerical results obtained with
its help are not  rigorous predictions of QCD in any well-defined limit.
The reason for this is that the prediction (\ref{qmff}) for the
form factors $f_{V,A}(E_\gamma)$ receives uncontrollable corrections
of order $\Lambda^2/E_\gamma^2$ as soon as the photon
energy $E_\gamma$ does not lie within the region of applicability of our analysis
$\Lambda_{QCD} \ll E_\gamma$. A similar statement can be made about
the corresponding predictions for the charged lepton energy spectrum, which
requires knowledge of the form factors over the entire range of $E_\gamma$.

In order to avoid these problems, we will restrict our considerations
to quantities defined with a sufficiently high lower cut on
$E_\gamma$.
When radiative corrections are taken into account, the
hadronic matrix element $R$ in (\ref{qmff}) acquires a logarithmic
dependence on $E_\gamma$ given by (\ref{scalingV})
\bea\label{R(E)}
R(E_\gamma) = \frac{1}{\langle (k_+)^0\rangle}
\left(\frac{\alpha_s(m_b)}{\alpha_s(2E_\gamma)}
\right)^{-2/\beta_0}
\int_0^\infty \mbox{d}k_+
\frac{\tilde \psi(k_+)}{k_+}\left(1 + \frac{\alpha_s C_F}{4\pi} \ell(E_\gamma)
\right) + {\cal O}\left(\frac{\Lambda}{E_\gamma},
\frac{\Lambda}{m_b}\right)\,.
\eea
We show in Fig.~3 the results obtained for the form factors and in
Fig.~4 for the photon energy spectrum using
the tree-level form factors and including the one-loop correction computed in
Section III.
This correction decreases the rate, at least in the region of
validity of our results. This effect is mostly due to the double log
in the one-loop hard scattering amplitude; the leading-log factor in
(\ref{scalingV}) makes a positive contribution. This illustrates the importance
of the double logarithms $\log^2(\frac{2E_\gamma}{k_+})$, which have to
be resummed to all orders.
The third curve in Figs.~3 and 4 shows the spectrum obtained by
resumming the Sudakov logarithms to all orders, as explained in Sec.~IV.

While the functional form of the hadronic matrix element $R(E_\gamma)$
depends on the detailed form of the (unknown) $B$ meson light-cone 
wave function, it is important to note that it is independent of the 
heavy quark mass $m_b$ (up to calculable logarithmic corrections). 
One would like to eliminate it by taking ratios of the photon spectra 
in $B$ and $D$ radiative leptonic decays. However, the large value of 
the $1/m_c$ correction in the latter case would introduce large 
corrections to such a ratio, which shows that some knowledge of
$R(E_\gamma)$ is necessary.

With this view in mind, we propose in the following a two-step 
procedure for determining the magnitude of the CKM matrix element 
$|V_{ub}|$. In the first step, the hadronic function
$R^{(b)}(E_\gamma)$
is determined in a region $E_\gamma \gg \Lambda_{QCD}$ from the 
normalized photon spectrum in $B^+$ decays
\bea\label{normalized_ratio}
\frac{1}{\Gamma_\ell(B^+\to \mu^+\nu)}
\frac{d\Gamma(B^+\to \gamma e^+\nu)}{dE_\gamma} =
\frac{\alpha m_B}{3\pi} \left(
Q_u R^{(b)}(E_\gamma) - \frac{Q_b}{m_b}\right)^2
\left(\frac{m_B}{m_\mu}\right)^2\frac{x_B(1-x_B)}
{1-\frac{m_\mu^2}{m_B^2}}\,,
\eea
with $x_B\equiv 1-(2E_\gamma)/m_B$. We used on the RHS the leading
twist result for the $B^+$ form factors; the $1/m_b$ correction is 
very small and will be neglected. The superscript on 
$R^{(b)}(E_\gamma)$ labels the heavy quark flavor.

In the second step, one takes the ratio of photon spectra in $B$ 
and $D$ decays, which is given by
\bea\label{B_to_Dratio}
\frac{\frac{d}{dE_\gamma}\Gamma(B^+\to \gamma e^+\nu)}
{\frac{d}{dE_\gamma}\Gamma(D^+\to \gamma e^+\nu)} &=&
\left|\frac{V_{ub}}{V_{cd}}\right|^2
\left(\frac{Q_u R^{(b)}(E_\gamma)}{Q_d R^{(c)}(E_\gamma) - Q_c/m_c}
\right)^2
\left(\frac{m_B}{m_{D}}\right)^3
\left(\frac{f_B}{f_{D}}\right)^2 \frac{x_B}{x_{D}} +
\cdots\\
&=& 
\left|\frac{V_{ub}}{V_{cd}}\right|^2
\left(\frac{Q_u R^{(b)}(E_\gamma)}{Q_d \zeta R^{(b)}(E_\gamma) 
- Q_c/m_c}\right)^2
\left(\frac{\alpha_s(m_b)}{\alpha_s(m_c)}\right)^{-4/\beta_0}
\left(\frac{m_B}{m_{D}}\right)^2
\frac{x_B}{x_{D}}\,,\nonumber
\eea
where $R^{(b)}(E_\gamma)$ is known from (\ref{normalized_ratio}).
We used here the logarithmic dependence on the heavy quark mass
(\ref{R(E)}) for the $R^{(Q)}(E_\gamma)$ coefficients
\beq
R^{(c)}(E_\gamma) = \zeta R^{(b)}(E_\gamma)\,,\qquad
\zeta = \left(\frac{\alpha_s(m_c)}{\alpha_s(m_b)}\right)^{-2/\beta_0}
\eeq
and the large mass scaling law \cite{VoSh,PoWi} for
the pseudoscalar decay constants
\bea
\frac{f_B}{f_{D}} \simeq
\left(\frac{\alpha_s(m_b)}{\alpha_s(m_c)}\right)^{-2/\beta_0}
\sqrt{\frac{m_{D}}{m_B}} \,.
\eea
The result (\ref{B_to_Dratio}) can be used to determine the 
CKM matrix element $|V_{ub}|$.

The leading corrections to this determination come from higher-twist
effects of order $O(\Lambda^2/E_\gamma^2)$ in the $D^+$ meson radiative 
leptonic form factors. Their magnitude can be estimated by comparing 
the normalized photon spectra (\ref{normalized_ratio}) in $B$ and $D$ 
decays.
Although for the $B$ case these corrections are expected to be well
under control over a reasonably wide range of values for $E_\gamma$,
it is questionable whether in the $D$ case such an
large energy region $\Lambda_{QCD} \ll E_\gamma$
exists at all. Since the maximum photon energy accessible in $D$
decays is only about $0.93$ GeV, the higher twist effects can be
expected to contribute no less than 10\% to the $D$ meson form factors.

A similar determination of $|V_{ub}|$ can be performed using
instead of $D^+$, the more accessible $D_s^+$ meson radiative 
leptonic decays. However, this would introduce an additional 
uncertainty on the theoretical side through SU(3) breaking effects.

\section{Conclusions}

We studied in this paper the form factors for the radiative
leptonic decay of a heavy meson (e.g. $B^+\to\gamma e^+\nu$) in an
expansion in powers of the inverse photon energy $1/E_\gamma$.
To leading order ${\cal O}(1/E_\gamma)$ these form factors are given 
by a convolution of the light-cone $B$ meson wave function 
$\psi(k_+)$ with an infrared-finite hard scattering kernel
$T_H(k_+,\vec k_\perp)$ (\ref{TH}).

Physically, this problem is very similar to the $\pi\gamma\gamma^*$
pion form factor $F_{\pi\gamma}(Q^2)$ studied in \cite{BL,Braaten},
where a similar factorization can be established for the leading
twist contribution of ${\cal O}(1/Q^2)$. However, there are some
important differences, the most striking of which concerns the
dependence on the transverse momentum to leading twist revealed in
the form of the hard scattering amplitude $T_H(k_+,\vec k_\perp)$.
Such a dependence is absent in the case of the pion form factor,
and its appearance can be traced to the presence of the
second dimensional parameter $k_+$ (the light-cone projection of
the light quark momentum in the $B$ meson) in addition to the large
scale $E_\gamma$. On the practical side, this implies a certain
loss of predictive power: while the logarithmic dependence on
$E_\gamma$ is well-determined, the constant term depends on the
precise form of the full 3-dimensional light-cone $B$ wave function.

A second important complication compared to the $F_{\pi\gamma}(Q^2)$
case consists in the appearance of Sudakov double logarithms,
which have to be resummed to all orders. This feature has been noted
previously in the context of the $B\to \pi(\rho)$ semileptonic
form factors of a heavy hadron in \cite{AkStYa,HNLi}, where these
Sudakov effects have been resummed
(up to next-to-leading order). Numerically, their effect
is most important near the upper end of the photon energy spectrum.

An interesting qualitative result of our analysis is the equality of
form factors of different currents $f_V(E_\gamma) = f_A(E_\gamma)$
at leading twist. While this equality was established by an explicit
one-loop calculation, it is probably a general result, true to
all orders in the strong coupling. In a perturbative QCD language,
the reason for this equality roots in the dominance of the momentum
integration regions where the propagator of the struck quark
(see Figs.~2) can be approximated with a light-like eikonal line.
This relation can be formalized by going over to an effective theory
\cite{DuGr} where the couplings of gluons to this line possess a
higher symmetry.

A similar approach has been taken in \cite{relations} to derive 
relations among semileptonic decay form factors of a heavy hadron. 
However, in the latter case the hard one-gluon exchange mechanism 
can be shown to introduce corrections to these relations, already
at leading twist. This is different from our case where these 
relations appear to be preserved (at least at one-loop order) under 
inclusion of the hard gluon exchange.

Finally, our formalism can be used to put previous quark model estimates
of radiative leptonic decays \cite{AES} on a more firm theoretical
basis, by giving a precise definition of the light quark constituent
mass. Our approach is likely to give a reliable description of the
form factors in the large $E_\gamma$ region, up to controllable corrections
of order $\Lambda^2/E_\gamma^2$. This complements an alternative approach
presented in \cite{BGW} which is best suited to the low-$E_\gamma$ region,
where the heavy hadron chiral perturbation theory is expected to be applicable.

Using as input parameter the binding energy of a $B$ meson
$\bar\Lambda$, we gave several estimates for the branching ratios
of these modes. As a by-product, we presented also a method for
extracting the  CKM matrix element $|V_{ub}|$ by
comparing photon energy spectra in radiative leptonic $B$ and $D$
decays.

\acknowledgements

D. P. is grateful to Peter Lepage for many discussions about the
application of perturbative QCD to exclusive processes, and to
Benjamin Grinstein for informing him of a related work \cite{GrRo}
before publication. This work has been supported in part by the
National Science Foundation.

\newpage
\appendix
\section{Scalar integrals}

We present here a few details relevant for the computation of the 
radiative corrections. The scalar integral appearing in the 
heavy-light
vertex correction $J^{(a)}$ is computed by first combining
the two massless propagators with the help of a Feynman parameter
\bea\nonumber
J^{(a)}&=&\int\frac{\mbox{d}^4 l}{(2\pi)^4}
\frac{1}{(-v\cdot l+i\epsilon)((l+\tilde q-k)^2+i\epsilon)
(l^2+i\epsilon)}\\
& &\qquad =
\int_0^1\mbox{d}x \int\frac{\mbox{d}^4 l}{(2\pi)^4}
\frac{1}{(-v\cdot l+i\epsilon)((l+x(\tilde q-k))^2 - s+i\epsilon)^2}
\nonumber\\
& &\qquad
= -\frac{i}{8\pi^2}\int_0^\infty \mbox{d}l_+
\int_0^1 \mbox{d}x
\frac{1}{l_+^2 - 2x\tilde q_0 l_+ + s}\,,
\eea
with $s = -x(1-x)(\tilde q-k)^2$\,.
After shifting the loop momentum $l\to l-x(\tilde q-k)$, one integrates
over the light-cone component $l_-$ using the Cauchy theorem, and
subsequently over the transverse momentum $l_\perp$.

The integral with one power of $l_\alpha$ in the numerator can be
reduced to a two-point function plus a UV finite integral by
first combining the massless denominators with a Feynman parameter
as above. This gives for the numerator
\bea
l_\alpha &\to & l_\alpha -x(\tilde q-k)_\alpha
= (v\cdot l) v_\alpha + (l_\perp)_\alpha - x(\tilde q-k)_\alpha \\
&=& -[-v\cdot l + xv\cdot (\tilde q-k)] v_\alpha
- x[(\tilde q-k)_\alpha - v\cdot (\tilde q-k)] +
(l_\perp)_\alpha\,.\nonumber
\eea
The first term cancels the heavy quark propagator in the
denominator, and the $(l_\perp)_\alpha$ term vanishes after
integration over $l$. One obtains in this way
\bea
J_\alpha^{(a)} &\equiv&
\int\frac{\mbox{d}^4 l}{(2\pi)^4}
\frac{l_\alpha}{(-v\cdot l+i\epsilon)((l+\tilde q-k)^2+i\epsilon)
(l^2+i\epsilon)}\\
&=& -v_\alpha \Sigma^{(a)} -
[(\tilde q-k)_\alpha - v\cdot (\tilde q-k)]\langle xJ^{(a)}\rangle
\eea
with
\bea
\Sigma^{(a)} = \int\frac{\mbox{d}^n l}{(2\pi)^n}
\frac{1}{[(l+\tilde q-k)^2+i\epsilon] (l^2 + i\epsilon)}
\eea
and
\bea
& &\langle xJ^{(a)}\rangle =
-\frac{i}{8\pi^2}\int_0^\infty \mbox{d}l_+
\int_0^1 \mbox{d}x
\frac{x}{l_+^2 - 2x\tilde q_0 l_+ + s}
\eea

These integrals can be evaluated exactly with the following results
\bea\label{Jexact}
J^{(a)} &=& -\frac{i}{(4\pi)^2}\frac{1}{(E_\gamma-\bar\Lambda)\sqrt{\xi}}
\left\{
-\mbox{Li}_2\left(-\frac{r_1\sqrt{\xi}-2E_\gamma k_+}{
2E_\gamma k_+(\sqrt{\xi}+1)}\right)
+\mbox{Li}_2\left(-\frac{r_2\sqrt{\xi}-2E_\gamma k_+}{
2E_\gamma k_+(\sqrt{\xi}+1)}\right)\right.\\
& &\hspace{4cm} -\left. 3\mbox{Li}_2\left(-\frac{\sqrt{\xi}-1}{\sqrt{\xi}+1}\right)
+ \frac{\pi^2}{2}
\right\}\nonumber\,,\\
\label{(1-x)J}
\langle (1-x)J^{(a)}\rangle &=& J^{(a)} - \langle xJ^{(a)}\rangle=
\frac12(1+\frac{1}{\xi})J +
\frac{i}{(4\pi)^2}\frac{1}{4(E_\gamma-\bar\Lambda)\xi}
\left\{
6\log\frac{4(E_\gamma-\bar\Lambda)^2\sqrt{\xi}}{2E_\gamma k_+}\right.\\
 &+& \left.(1+\frac{r_1}{2E_\gamma k_+})
\log\frac{8(E_\gamma-\bar\Lambda)^2\sqrt{\xi}}{2E_\gamma k_++r_1}
- (1+\frac{r_2}{2E_\gamma k_+})
\log\frac{8(E_\gamma-\bar\Lambda)^2\sqrt{\xi}}{2E_\gamma k_++r_2}\right.\nonumber\\
&-& \left.
6\log\frac{2\sqrt{\xi}}{1+\sqrt{\xi}} -
\frac{2E_\gamma k_+}{(E_\gamma-\bar\Lambda)^2(1+\xi)}
\log\frac{2E_\gamma k_++r_1}{2E_\gamma k_++r_2}\right.\nonumber\\
&+& \left.
\frac{(2E_\gamma k_+)^2}{(E_\gamma-\bar\Lambda)^2(2E_\gamma k_+-r_1)\xi}
\log\frac{2E_\gamma k_++r_1}{2E_\gamma k_+(1+1/\sqrt{\xi})}\right.\nonumber\\
&-& \left.
\frac{(2E_\gamma k_+)^2}{(E_\gamma-\bar\Lambda)^2(2E_\gamma k_+-r_2)\xi}
\log \frac{2E_\gamma k_++r_2}{2E_\gamma k_+(1+1/\sqrt{\xi})}
 \right\}\nonumber
\eea
with $\xi=1+2E_\gamma k_+/(E_\gamma-\bar\Lambda)^2$ and
$r_{1,2} = 4(E_\gamma-\bar\Lambda)^2 (\pm 1 + \sqrt{\xi}) \pm 2E_\gamma k_+$.

The vertex correction to the photon coupling to the light quark is
parametrized in terms of the integrals
\bea\label{J(b)}
J^{(b)}&=& \frac{i}{(4\pi)^2}\frac{1}{2p_\gamma\cdot k}
\left( -\frac12\log^2\frac{2p_\gamma\cdot k}{m^2} -
\frac{\pi^2}{3}\right)\\
\nonumber
J^{(b)}_1 &=& \frac{i}{(4\pi)^2}\frac{1}{2p_\gamma\cdot k}
\left( \log\frac{2p_\gamma\cdot k}{m^2}
-\frac12\log^2\frac{2p_\gamma\cdot k}{m^2} - \frac{\pi^2}{3} - 1
\right)\\
\nonumber
J^{(b)}_2 &=& \frac{i}{(4\pi)^2}\frac{1}{2p_\gamma\cdot k}
\left( \log\frac{2p_\gamma\cdot k}{m^2} - 2\right)\,,\quad
J^{(b)}_3 = \frac{i}{4(4\pi)^2}\left( N_\varepsilon^{UV} + 3 -
\log\frac{2p_\gamma\cdot k}{\mu_{\rm h}^2}
\right)\\
\nonumber
J^{(b)}_5 &=& \frac{i}{(4\pi)^2}\frac{1}{2p_\gamma\cdot k}
\left( \log\frac{2p_\gamma\cdot k}{m^2} - \frac52\right)
\,.
\eea

When computing the one-loop correction to $f_B$ and the box diagram, one
encounters the IR singular integral
\bea
J_{IR}(v\cdot k) &=& \mu^{2\epsilon}\int\frac{d^D l}{(2\pi)^D}
\frac{1}{(-v\cdot l+i\varepsilon)(l^2+2l\cdot k+i\varepsilon)
(l^2+i\varepsilon)}\\
&=& \frac{i}{(4\pi)^2}\frac{1}{\epsilon}\Gamma(1+\epsilon)
\int_0^\infty
dx \frac{(4\pi\mu^2 x^2)^{\epsilon}}
{[(xk-v)^2-i\varepsilon]^{1+\epsilon}}\nonumber\,.
\eea
The IR singularity has been regulated with dimensional regularization
in $D=4-2\epsilon$ dimensions. The integral over $x$ can be computed
explicitly with the result  (with $Q=v\cdot k$)
\bea\label{JIR(Q)}
J_{IR}(Q) &=&
\frac{i}{(4\pi)^2} \frac{1}{2\sqrt{Q^2-m^2}}
\left\{
(N_\varepsilon^{IR} + \log\frac{\mu^2}{m^2}) \left[
\log\frac{Q-\sqrt{Q^2-m^2}}{Q+\sqrt{Q^2-m^2}} + 2\pi i\right]
\right.\\
&+&\left.
\mbox{Li}_2\left(
\frac{2\sqrt{Q^2-m^2}}{Q+\sqrt{Q^2-m^2}}\right)
-\mbox{Li}_2\left(
\frac{-2\sqrt{Q^2-m^2}}{Q-\sqrt{Q^2-m^2}} \right) +
\frac{\pi^2}{6} - 2\pi i\log\frac{4(Q^2-m^2)}{m^2}
\right\}\nonumber
\end{eqnarray}
with $N_\varepsilon^{IR} = 1/\epsilon - \gamma_E + \log(4\pi)$.
In the limit $Q \gg m$ this agrees with the expression given in the
Appendix C of \cite{BeDe}.

\begin{figure}[hhh]
 \begin{center}
 \mbox{\epsfig{file=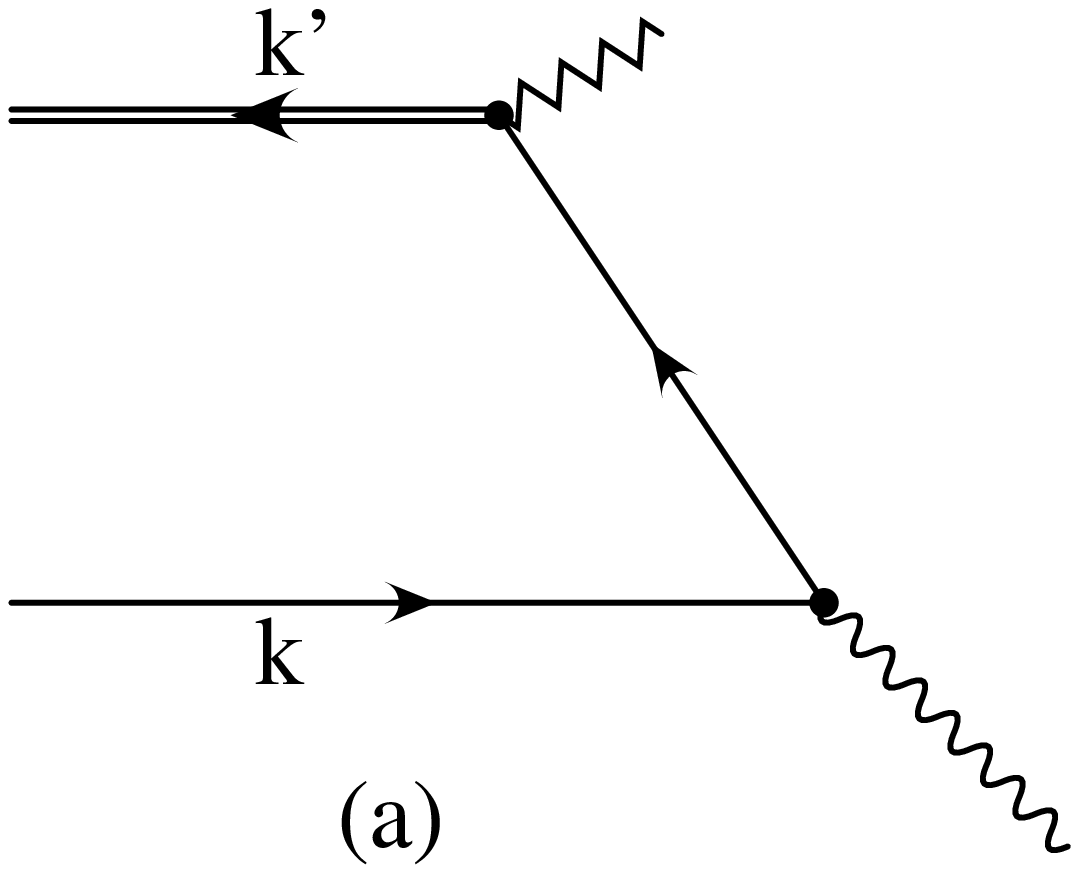,width=4cm}\qquad\qquad
 \epsfig{file=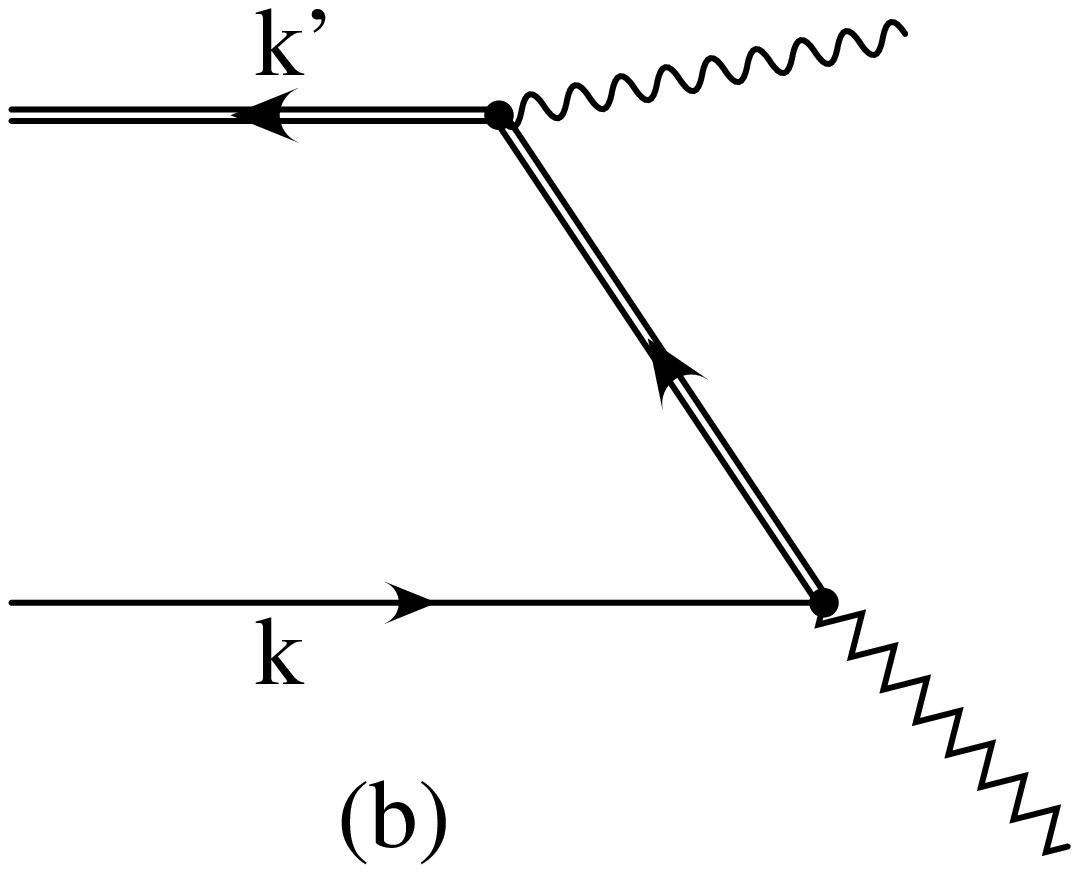,width=4cm}}
 \end{center}
 \caption{
Leading order diagrams contributing to the radiative leptonic decay
$B^+\to W\gamma$. The double line denotes the heavy quark $b$, the
zigzag line the W boson and the wiggly line a photon.}
\label{fig1}
\end{figure}

\begin{figure}[hhh]
 \begin{center}
 \mbox{\epsfig{file=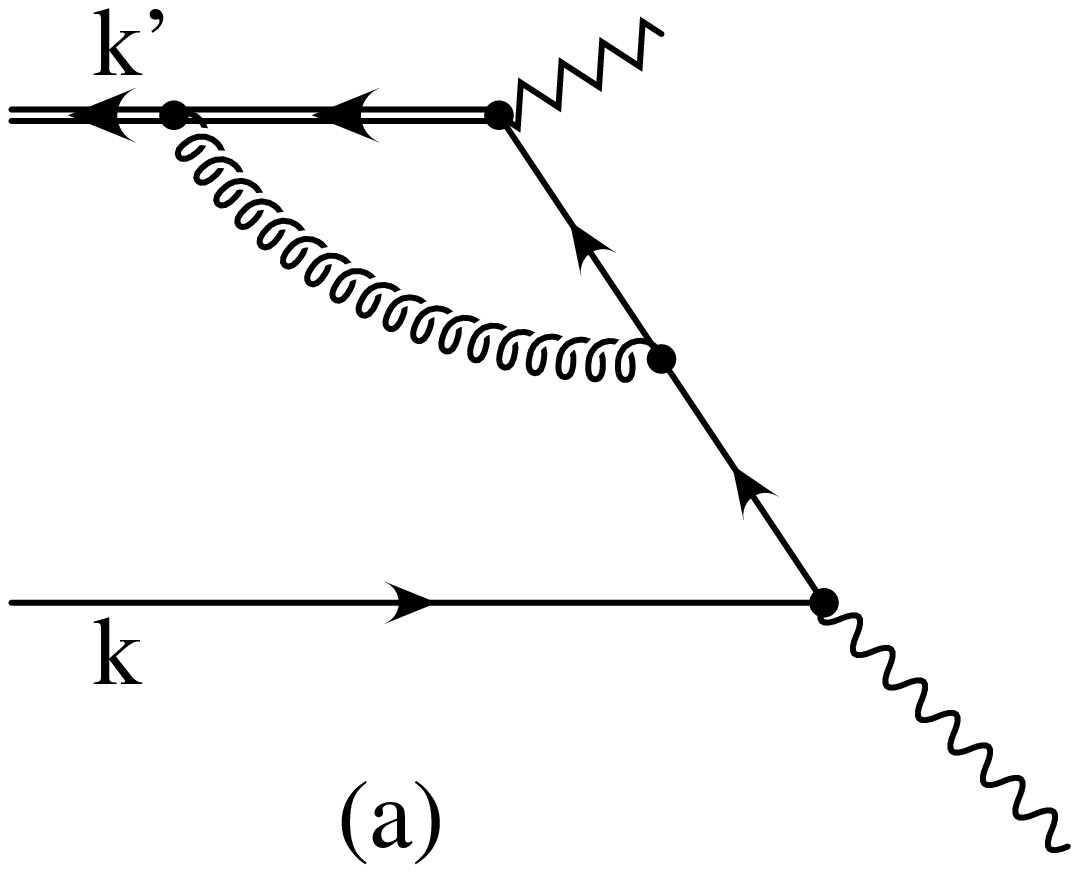,width=4cm}\qquad\qquad
\epsfig{file=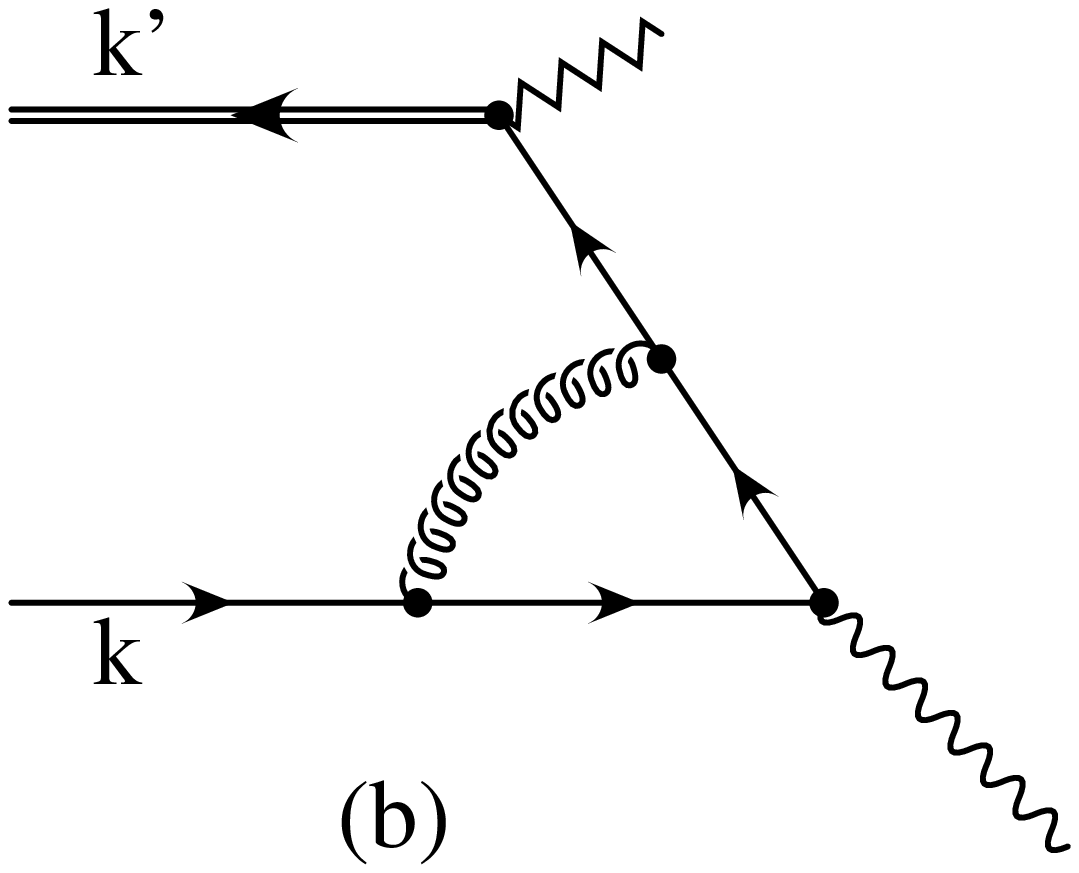,width=4cm}}
 \mbox{\epsfig{file=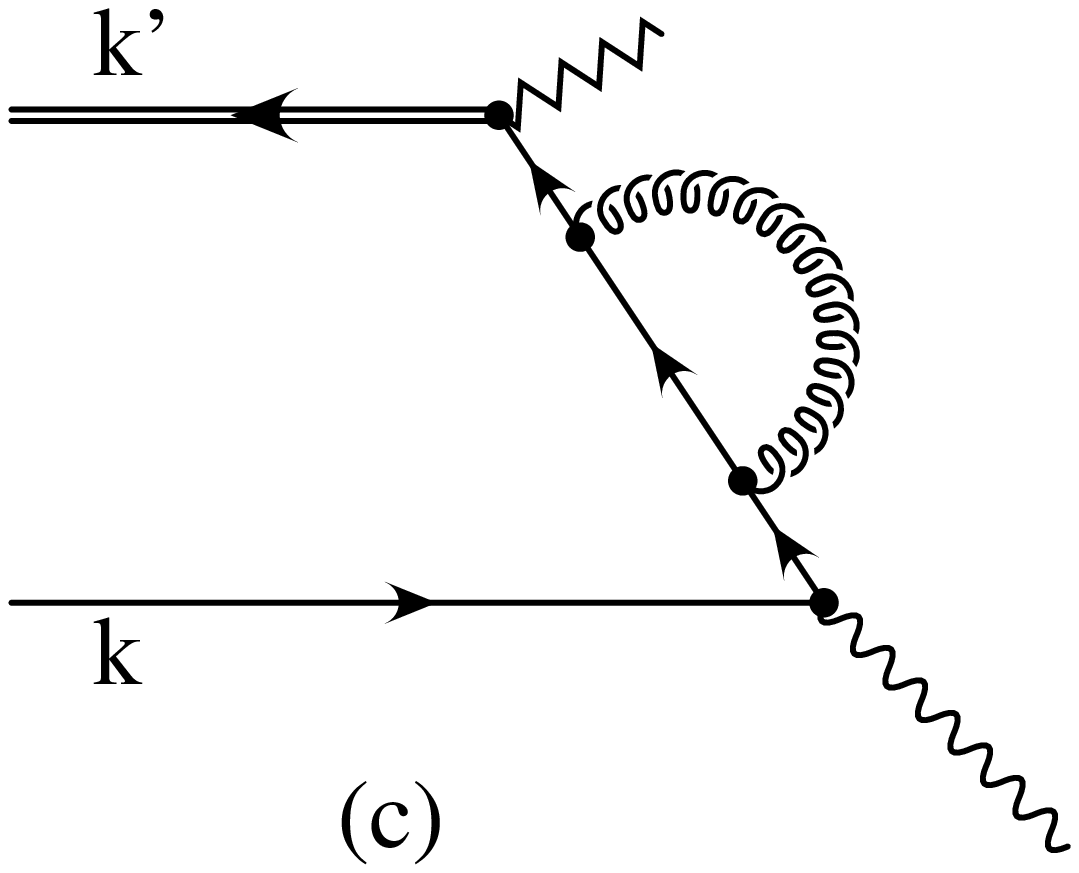,width=4cm}\qquad\qquad
\epsfig{file=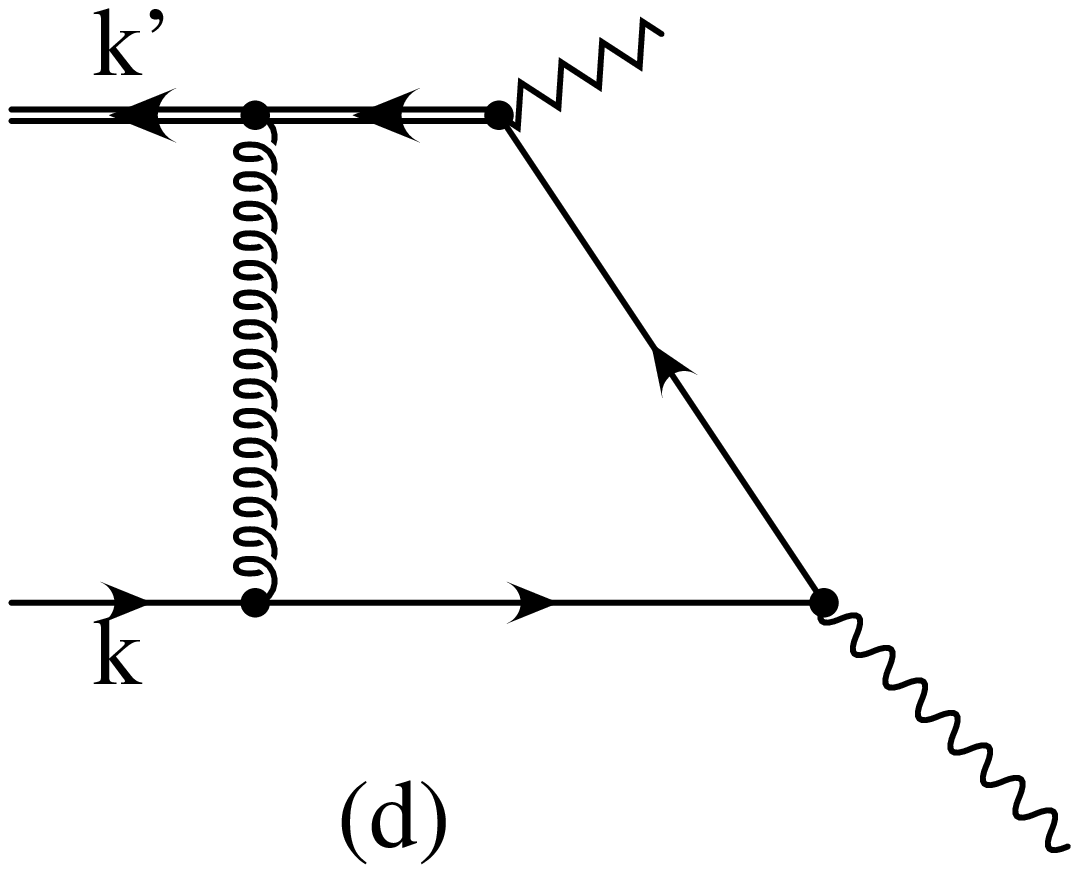,width=4cm}}
 \end{center}
 \caption{
One-loop corrections to the radiative leptonic decay
$B^+\to W\gamma$. The curly line represents a gluon.
The quark wave function renormalization corrections are not shown.}
\label{fig2}
\end{figure}

\newpage

\begin{figure}[hhh]
 \begin{center}
 \mbox{\epsfig{file=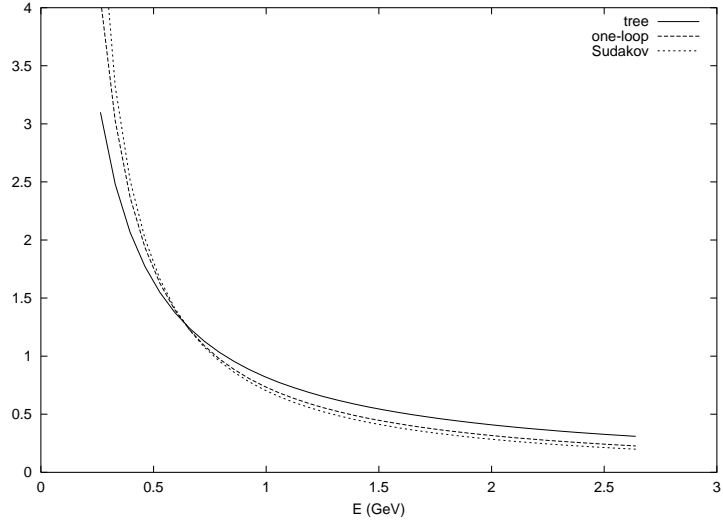,width=10cm}}
 \end{center}
\caption{
Typical leading twist form factors $f_i(E_\gamma)$ $(i=V,A)$ for
$B\to \gamma e\nu$ decays. The continuous line shows the tree-level
result, the dotted line includes one-loop corrections to the
hard scattering amplitude, and the dashed line includes the resummed
Sudakov form factor truncated with a cut-off at $(k_+)_{min}=
\Lambda_{QCD}$. We use $\alpha_s(m_b)=0.3$ and $a=0.36$ GeV,
$\omega=0.2$ GeV, corresponding to $\bar\Lambda=0.35$ GeV.}
\label{fig3}
\end{figure}

\begin{figure}[hhh]
 \begin{center}
 \mbox{\epsfig{file=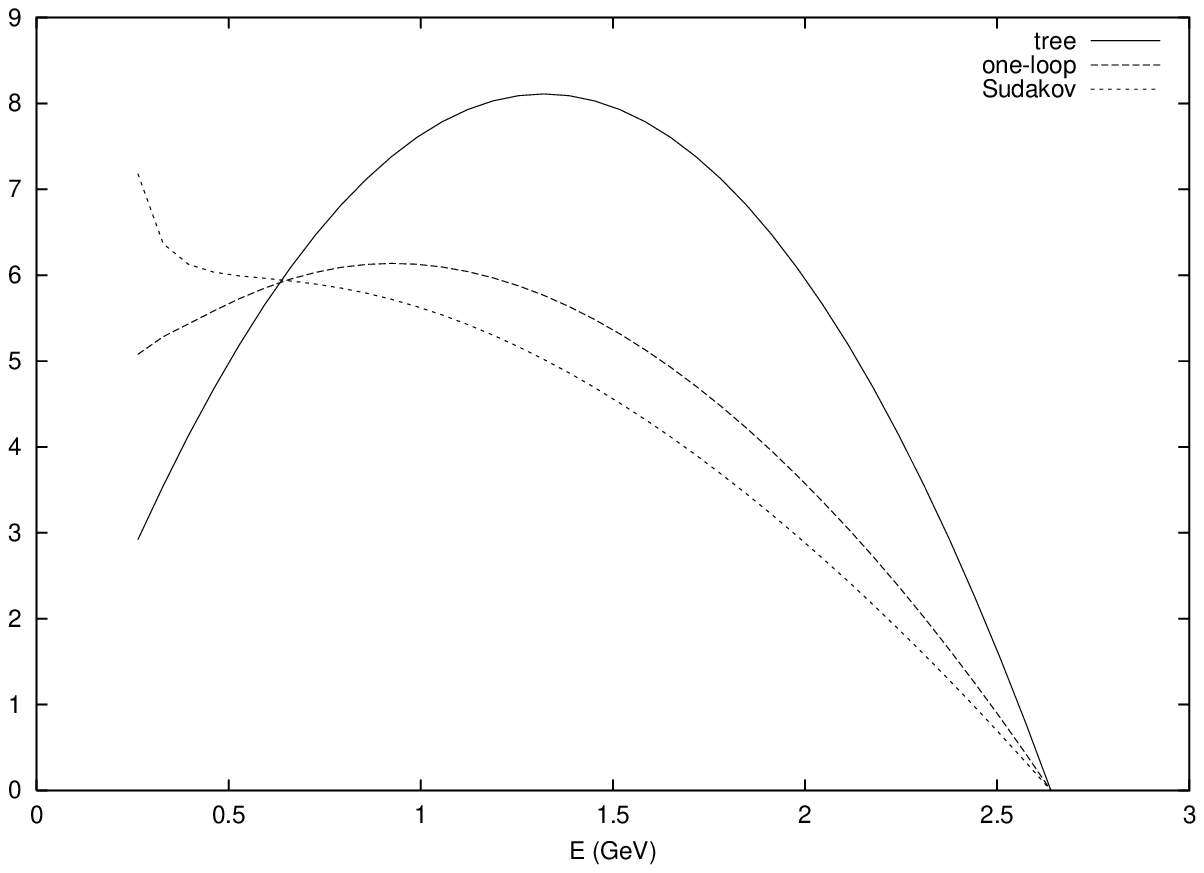,width=10cm}}
 \end{center}
\caption{The photon spectrum in $B^+\to\gamma e^+\nu$ normalized
to the pure muonic leptonic decay rate. The continuous line represents
the tree-level result, assuming $R=3$ GeV$^{-1}$. The dotted line includes
the effects of the one-loop strong correction (only the logarithms) with
$\alpha_s(m_b)=0.3$, and the dashed line includes the resummed Sudakov
logs. The same parameters are used for the light-cone $B$ wave
function as in Fig.~3.}
\label{fig4}
\end{figure}


\begin{thebibliography}{99}
\bibitem{BGW} G. Burdman, T. Goldman and D. Wyler, Phys. Rev. D{\bf 51}, 111 (1995).

\bibitem{Wyler} A. Khodjamirian, G. Stoll and D. Wyler,
   Phys. Lett. {\bf B358} (1995) 129.

\bibitem{CdFN1} P. Colangelo, F. De Fazio and G. Nardulli,
      Phys. Lett. B{\bf 372}, 331 (1996).

\bibitem{CdFN} P. Colangelo, F. De Fazio and G. Nardulli, Phys. Lett. B{\bf 386}, 328 (1996).

\bibitem{AES} D. Atwood, G. Eilam and A. Soni, Mod. Phys. Lett. A{\bf 11}, 1061 (1996).

\bibitem{LCSR} G. Eilam, I. Halperin and R. R. Mendel, Phys. Lett. B{\bf 361}, 137 (1995).

\bibitem{LFM} C.Q. Geng, C.C. Lih and Wei-Min Zhang, Phys. Rev. D{\bf 57}, 5697 (1998).

\bibitem{beta} J. F. Amundson {\em et al.}, Phys. Lett. B{\bf 296},
   415 (1992); P. Cho and H. Georgi, Phys. Lett. B{\bf 296}, 408
   (1992), Erratum-ibid. B{\bf 300}, 410 (1993); T. M. Yan {\em et al.}, Phys. Rev. D{\bf 46}, 1148 (1992).

\bibitem{dilept} G. Eilam, C. D. L\"u and D. X. Zhang,  Phys. Lett. B{\bf 391}, 461 (1997).

\bibitem{CLEOrad} T.E. Browder {\em et al.} (CLEO Collaboration),
   Phys. Rev. D{\bf 56} (1997) 11.

\bibitem{BL} S. Brodsky and P. Lepage, Phys. Rev. D{\bf 22}, 2157 (1980).

\bibitem{pQCD} S. Brodsky and P. Lepage, in {\em Perturbative Quantum
Chromodynamics}, by A. H. Mueller (ed.), World Scientific 1989.

\bibitem{GrNe} A. Grozin and M. Neubert, Phys. Rev. D{\bf 55}, 
   272 (1997).

\bibitem{VoSh} M. A. Shifman and M. B. Voloshin,
   Sov. J. Nucl. Phys. {\bf 45} (1987) 292; {\bf 47} (1988) 511.

\bibitem{PoWi} H. D. Politzer and M. B. Wise,
   Phys. Lett. {\bf B206} (1988) 681; {\bf B208} (1988) 504.

\bibitem{relations} J. Charles {\em et al.}, Phys. Rev.
   {\bf D60} (1999) 14001; J. Charles {\em et al.}, Phys. Lett.
   {\bf B451} (1999) 187.

\bibitem{DuGr} M. J. Dugan and B. Grinstein, Phys. Lett.
   {\bf B255} (1991) 583.

\bibitem{Braaten}
  F. del Aguila and M. K. Chase, Nucl. Phys. {\bf B193} (1981) 517;
  E. Braaten, Phys.Rev. {\bf D28} (1983) 524; E. P. Kadantseva,
  S. V. Mikhailov and A. V. Radyushkin, Yad.Fiz. {\bf 44}
  (1986) 507 [Sov. J. Nucl. Phys. {\bf 44} (1986) 326].

\bibitem{GKPLB} G. P. Korchemsky, Phys. Lett. {\bf B220} (1989) 629.

\bibitem{AkStYa} R. Akhouri, G. Sterman and Y. P. Yao,
   Phys. Rev. {\bf D50} (1994) 358.

\bibitem{HNLi} H. N. Li and H. L. Yu, Phys. Rev. {\bf D53} (1996) 2480.

\bibitem{CoSo} A. H. Mueller, Phys. Rev. {\bf D20} (1979) 2037;
   J. C. Collins, Phys. Rev. {\bf D22} (1980) 1478;
   G. P. Korchemsky and A. V. Radyushkin, Yad. Fiz. {\bf 45} (1987)
   1460 [Sov. J. Nucl. Phys. {\bf 45} (1987) 910.]

\bibitem{KoMa} G. Korchemsky and G. Marchesini,
   Phys. Lett. {\bf B313} (1993) 433;
   Nucl. Phys. {\bf B406} (1993) 225.

\bibitem{BSW} M. Bauer and M. Wirbel, Z. Phys. {\bf C42} (1989) 671.

\bibitem{BSS} M. Bander, D. Silverman and A. Soni, Phys. Rev. Lett.
   {\bf 44} (1980) 7; (E) {\bf 44} (1980) 962.

\bibitem{AtBlSo} D. Atwood, B. Blok and A. Soni, Int. J. Mod. 
   Phys. {\bf A11} (1996) 3743.

\bibitem{Cheng} H. Y. Cheng {\em et al}, Phys. Rev. {\bf D51} (1995) 
   1199.

\bibitem{BBNS} M. Beneke {\em et al.}, Phys. Rev. Lett. {\bf 83}
   (1999) 1914.

\bibitem{Vub} B. H. Behrens {\em et al.}, CLEO Collaboration,
   hep-ex/9905056.

\bibitem{Lig} For a recent review of various $|V_{ub}|$ determinations see
 Z. Ligeti, hep-ph/9908432.

\bibitem{CLEO} M. Chadha {\em et al.}, Phys. Rev. {\bf D58} (1998) 032002-1.


\bibitem{BeDe} W. Beenakker and A. Denner,  Nucl. Phys. B{\bf 338} (1990)
   349.

\bibitem{GrRo} B. Grinstein and I. Rothstein, to be published
\end{thebibliography}
\end{document}